\documentclass[12pt,a4paper,dvips]{article}
\usepackage{a4p}
\usepackage{cite,mcite}
\usepackage{graphicx}
\usepackage{rotating,a4p,here}
\usepackage{phy_susy, amssymb}
\usepackage{amsmath}
\usepackage{l3_title,ifthen}
 \journalname{Eur. Phys. J.}
 \date{October 10, 2000}
 \preprint{2000-130}
\newlength{\capindent}
\newlength{\capwidth}
\newlength{\figwidth}
\newcommand{\icaption}[2][!*!,!]{\hspace*{\capindent}%
  \begin{minipage}{\capwidth}
    \ifthenelse{\equal{#1}{!*!,!}}%
      {\caption{#2}}%
      {\caption[#1]{#2}}
  \end{minipage}}
\newcommand{\pho}{\phantom{0}}
\newcommand{\phouno}{\phantom{1}}
\newcommand{\phoescl}{\phantom{!}}
\newcommand{\rpvpha}{\vphantom{$\overline{\slep_j}$}} 
\newcommand{\rpvpham}{\vphantom{$\underline{\mchid}$}}

\def\lamtre{\ifmath{\lambda_{133}}}
\def\lamuno{\ifmath{\lambda''_{212}}}
\def\ytq{\ifmath{y_{34}}}

\def\yqc{\ifmath{y_{45}}}

\def\ycut{\ifmath{y_{cut}}}

\def\Emiss{\ensuremath{E\hspace{-.23cm}/\hspace{+.01cm}}}
\def\LLE{\ifmath{\mathrm{L}_{i}\mathrm{L}_{j}{\overline{\mathrm E}_{k}}}}
\def\LQD{\ifmath{\mathrm{L}_{i}\mathrm{Q}_{j}{\overline{\mathrm D}_{k}}}}
\def\UDD{\ifmath{{\overline{\mathrm U}_{i}}{\overline{\mathrm D}_{j}}{\overline{\mathrm D}_{k}}}}
\def\snu{\ifmath{\tilde{\nu}}}

\def\slep{\ifmath{\tilde{\ell}}}
\def\slepp{\ifmath{\tilde{\ell}^+_R}}
\def\slepm{\ifmath{\tilde{\ell}^-_R}}
\def\slepr{\ifmath{\tilde{\ell}_R}}
\def\serr{\ifmath{\tilde{\rm e}_R}}
\def\serp{\ifmath{\tilde{\rm e}^+_R}}
\def\serm{\ifmath{\tilde{\rm e}^-_R}}
\def\smur{\ifmath{\tilde{\mu}_R}}
\def\smurp{\ifmath{\tilde{\mu}^+_R}}
\def\smurm{\ifmath{\tilde{\mu}^-_R}}
\def\staur{\ifmath{\tilde{\tau}_R}}
\def\staurp{\ifmath{\tilde{\tau}^+_R}}
\def\staurm{\ifmath{\tilde{\tau}^-_R}}
\def\susy#1{\ensuremath{\tilde{\mathrm{#1}}}}%

\def\neutralino#1{\ensuremath{\susy{\chi}_{#1}^0}}

\def\mo{\ensuremath{m_0~}}

\def\tb{\ensuremath{\tan\beta}}

\begin{document}
\begin{titlepage}
\title{ \boldmath{Search for R-parity Violating Decays of Supersymmetric
       Particles in $\mathrm{e^+e^-}$ Collisions at  
       $\sqrt{s}=$ 189~\GeV}}

\author{The L3 Collaboration}
%
%
\begin{abstract}
A search for chargino, neutralino and scalar lepton
pair-production in $\mathrm \epem$
collisions at the centre-of-mass energy of 189 \GeV{}
is performed under the assumptions that R-parity is not
conserved in decays and only one of the coupling constants 
$\lambda_{ ijk}$, $\lambda'_{ ijk}$ or $\lambda''_{ ijk}$
is non-negligible.
No signal is found in a data sample corresponding to an
integrated luminosity of 176.4 \pbi. 
Limits on the production cross sections, 
on the Minimal Supersymmetric Standard Model parameters
and on the masses of the supersymmetric particles  
are derived.

\end{abstract}

\submitted
\end{titlepage}

\section{Introduction}

The most general superpotential of the
Minimal Supersymmetric Standard Model (MSSM) ~\cite{MSSM}, which describes
a supersymmetric, renormalizable
and gauge invariant theory, with minimal particle content, includes
the term $\mathrm{W_R}$~\cite{superp,superp2}:
\begin{equation} \mathrm{W_R} =
    \lambda_{ ijk} \mathrm L_i \mathrm L_j \overline{\mathrm E}_k \, +
    \lambda'_{ ijk} \mathrm L_i \mathrm Q_j \overline{\mathrm D}_k \, +
    \lambda''_{ ijk} \overline{\mathrm U}_i \overline{\mathrm D}_j 
                            \overline{\mathrm D}_k\,,
\label{eqn:wr}
\end{equation}
where $\lambda_{ ijk}$, $\lambda'_{ ijk}$ and $\lambda''_{ ijk}$ 
are the Yukawa couplings
and {\it i, j} and {\it k} the generation indices;
$\mathrm L_i$ and $\mathrm Q_i$ are the left-handed lepton- and 
quark-doublet superfields,
$\overline{\mathrm E}_i$, $\overline{\mathrm D}_i$ and $\overline{\mathrm U}_i$
are the right-handed singlet superfields for
charged leptons, down- and up-type quarks, respectively.
In order to prevent the simultaneous presence of identical fermionic
fields, the following antisymmetry relations are required: 
$\mathrm \lambda_{ ijk} = - \lambda_{ jik}$
and $\mathrm \lambda''_{ ijk} = - \lambda''_{ ikj}$, 
reducing to  $\mathrm{9 + 27 + 9}$ the total number of independent 
Yukawa couplings.
The $\LLE$ and $\LQD$ terms violate the leptonic quantum number L, while 
the $\UDD$ terms violate the baryonic quantum number B. 

The simultaneous presence of L- and B-violating terms would 
lead to a fast proton decay\footnote{With contributions at the tree level 
from $\lambda'_{11k} \lambda''_{11k}$, and at one-loop level from
any product  $\lambda_{ijk} \lambda''_{lmn}$ or 
$\lambda'_{ijk} \lambda''_{lmn}$.}~\cite{weinberg}. Therefore, existing
limits~\cite{pdg2000} on the proton lifetime require either of two
possibilities. 

The first one is to impose R-parity conservation,
which forbids all terms in Equation~\ref{eqn:wr}. R-parity is
a multiplicative quantum number defined as:
\begin{equation} \mathrm
    R = (-1)^{\mathrm {3B+L+2S}}\,,
\end{equation}
where~S~is~the~spin.~R~ is $+1$~for~all~ordinary
particles,~and~$-1$~for~their~supersymmetric partners.
If R-parity is conserved, supersymmetric particles can be
produced only in pairs and they decay in cascade to the lightest supersymmetric
particle (LSP), which is stable~\cite{rpc_susy}.

The second possibility is considered in this paper. 
Since the absence of either the B-violating or the L-violating terms
is enough to prevent a fast proton decay, there is no need to
impose {\it a priori} R-parity conservation.
As a consequence, two new kinds of processes are allowed:
single production of supersymmetric particles, or
LSP decays into Standard Model particles
via scalar lepton or quark exchange. In the latter case, the MSSM
production mechanisms are unaltered by the operators in 
Equation~\ref{eqn:wr}. 

In this paper, we describe the
search for pair-produced neutralinos\footnote{Single production of
supersymmetric particles is not considered in this paper.}
($\mathrm \epem \ra \chim\chin$,
with $m = 1,2 \,$ and $n = 1,..,4$), 
charginos ($\mathrm \epem \ra \chap\cham$) 
and scalar leptons ($\mathrm \epem \ra \slepp \slepm$, 
$\epem \ra \snu\snu$)
with subsequent R-parity violating decays, assuming that only
one of the coupling constants $\lambda_{ ijk}$, $\lambda'_{ ijk}$ 
or $\lambda''_{ ijk}$ is non-negligible.
Only $\slepr$ (supersymmetric partners of the right-handed charged leptons)
are considered in this analysis, since they are expected to be lighter
than the corresponding left-handed ones.

Supersymmetric particles can decay directly into two or three fermions 
according to the dominant interaction term, as detailed in 
Table~\ref{tab:decays}. Indirect decays via the LSP can occur as well.
Four-body decays of the lightest scalar lepton are also taken into
account in the case of $\lambda'_{ ijk}$ and $\lambda''_{ ijk}$.
In the present analysis, the dominant coupling is assumed to be greater than
$10^{-5}$~\cite{dawson}, corresponding to decay lengths less than 1 cm.
Previous L3 results on $\lambda_{ ijk}$ and 
$\lambda''_{ ijk}$ Yukawa couplings can be found in Reference~\citen{rpv_l3}. 
Searches for R-parity violating decays of supersymmetric particles
were also performed by other LEP experiments~\cite{rpv_publ}.

\begin{table*} [htbp] 
  \begin{center}
  \begin{tabular}{|l|c|c|c|c|c|} \hline 

   \multicolumn{1}{|c|} {Particle}
   &\multicolumn{3}{c|} {Direct decays }
   &\multicolumn{2}{|c|} {Indirect decays } \\ \cline{2-6}

   \multicolumn{1}{|c|} { }
   &\multicolumn{1}{c|} {$\lambda_{ ijk}$}
   &\multicolumn{1}{c|} {$\lambda'_{ ijk}$}
   &\multicolumn{1}{c|} {$\lambda''_{ ijk}$}
   &\multicolumn{1}{|c|} { via $\neutralino{1}$}
   &\multicolumn{1}{|c|} { via \vphantom{$\overline{\slep_j}$} 
                          $\slep$} \\ \hline

     $\chio$  \vphantom{$\overline{\slep_j}$}
     & $ \ell_i^- \nu_j \ell_k^+$, $\nu_i \ell^+_j \ell^-_k$  
     & $  \ell_i^- {\mathrm u_j} \bar{\mathrm d}_k$, $\nu_i \, \mathrm d_j 
          \bar{\mathrm d}_k$  
     & $ \bar{\mathrm u}_i \bar{\mathrm d}_j \bar{\mathrm d}_k$  
     &  $-$   
     & $\ell\slep$ \\ \hline

    $\neutralino{n (n \ge 2)}$   \vphantom{$\overline{\slep_j}$}
     & $ \ell_i^- \nu_j \ell_k^+$, $\nu_i \ell^+_j \ell^-_k$  
     & $  \ell_i^- {\mathrm u_j} \bar{\mathrm d}_k$, $\nu_i \, \mathrm d_j 
          \bar{\mathrm d}_k$  
     & $ \bar{\mathrm u}_i \bar{\mathrm d}_j \bar{\mathrm d}_k$  
     &  $\Zstar\chio$, $\Zstar\chim_{(m < n)}$, 
     & $\ell\slep$ \\ 

      {  }
     & 
     & 
     & 
     & $\Wstar\cha$  
     & \\ \hline

    $\chap$   \vphantom{$\overline{\slep}_j$}
     & $ \nu_i \nu_j \ell^+_k$, $\ell^+_i \ell^+_j \ell^-_k$
     & $\nu_i u_j \bar{\mathrm d}_k$, $\ell^+_i  \bar{\mathrm d}_j 
        \mathrm d_k$ 
     & $\bar{\mathrm d}_i \bar{\mathrm d}_j \bar{\mathrm d}_k$, 
       $\mathrm u_i \mathrm u_j \mathrm d_k$,
     & $\Wstar\chio$, $\Wstar\chid$
     &  \\
 
        { }
     & 
     & 
     & $\mathrm u_i \mathrm d_j \mathrm u_k$ 
     & 
     &  \\ \hline

     $\slep^-_{kR}$   \vphantom{$\overline{\slep_j}$}
     & $\nu_i \ell^-_j$, $\nu_j \ell^-_i$ 
     & $-$
     & $-$
     & $\ell^-_k \chio$
     & $-$ \\ \hline

     $\snu_{i}$, $\snu_{j}$       
     & $\ell^-_j \ell^+_k$, $\ell^-_i \ell^+_k$
     & $-$
     & $-$
     & $\nu_i \chio$, $\nu_j \chio$
     &  \\  \hline

  \end{tabular}
  \caption{R-parity violating decays of the supersymmetric
   particles considered in this analysis.
   Charged conjugate states are implied. Indirect decays
   via scalar leptons are relevant only for neutralinos when the scalar
   lepton is the LSP. Only 
   supersymmetric partners of the right-handed charged leptons are taken
   into account. Decays to more than three fermions are not listed.
   $\Zstar$ and $\Wstar$ indicate virtual Z and W.}
  \label{tab:decays}
  \end{center}
\end{table*}

\section{Data and Monte Carlo Samples}

The data used  correspond to an integrated luminosity of 176.4 \pbi{} 
collected by the L3 detector~\cite{L3}
at the centre-of-mass energy ($\sqrt s$) of 189 \GeV.

The signal events are generated with the program {\tt SUSYGEN}~\cite{susygen}
for different mass values and
for all possible choices of the generation indices.

The following Monte Carlo
generators are used to simulate Standard Model background processes: 
{\tt PYTHIA}~\cite{pythia} for 
$\mathrm \epem \ra \qqbar$, 
$\; \mathrm \epem \ra \mathrm Z \, \epem$ and $\mathrm \epem \ra$ ZZ,
{\tt BHWIDE}~\cite{bhwide} for $\mathrm \epem \ra \epem $,
{\tt KORALZ}~\cite{koralz} for $\mathrm \epem \ra \mpmm$ and
$\mathrm \epem \ra \tautau$,
{\tt PHOJET}~\cite{phojet} and {\tt PYTHIA}  
for $\mathrm \epem \ra \epem$ hadrons,
{\tt DIAG36}~\cite{diag36} for $\mathrm \epem \ra 
\epem \ell^+ \ell^-$ ($\mathrm{\ell = e, \mu, \tau}$),  
{\tt KORALW}~\cite{koralw} for $\mathrm \epem \ra \WpWm $ and
{\tt EXCALIBUR}~\cite{EXCALIBUR} for
$\ee \rightarrow \mathrm{W^\pm\, \ell\,^\mp \nu}$.
The number of simulated events corresponds to 
at least 50 times the luminosity of the data,
except for Bhabha and two-photon processes, 
where the Monte Carlo samples correspond to 
2 to 6 times the luminosity.

The detector response is simulated using the {\tt GEANT} 
package~\cite{geant}. It takes into account effects of energy loss,
multiple scattering and showering in the detector materials. Hadronic 
interactions are simulated with the
{\tt GHEISHA} program~\cite{gheisha}. Time dependent inefficiencies
of the different subdetectors are also taken into account
in the simulation procedure.

\section{Event reconstruction}

Leptons ($\ell = $ e, $\mu, \tau$) and hadronic jets are reconstructed
as follows.
An electron is defined as an electromagnetic shower, with energy
greater than 1 \gev, matched with a track in the central chamber. 
Muon identification requires a track in the muon chambers matched with 
a track in the central chamber.
Hadronic tau decays are reconstructed from narrow isolated hadronic 
jets with energy greater than 2 \gev{} and one to three associated 
tracks. The tau energy must be contained in a cone of $10^\circ$ 
half-opening angle around the jet direction.
No additional tracks and no more than two additional calorimetric 
clusters are required in a further cone of $30^\circ$  half-opening 
angle. The ratio of the energies in the two cones has to be less than 2.0. 
Remaining clusters and tracks are classified as hadrons. Jets are 
reconstructed with the DURHAM algorithm~\cite{durham}.
The jet resolution parameter $y_{mn}$ is defined
as the $\ycut$ value at which the event configuration changes from $n$ to 
$m$ jets. 
$N_{jets8}$ is the number of jets clustered with $\ycut$ fixed 
and equal to 0.008. 
The acollinearity ($\theta_{acol}$) 
and acoplanarity ($\theta_{acop}$) angles are calculated 
by forcing hadronic and leptonic objects in every event into 
exactly two jets. At least one time of flight measurement 
has to be consistent with the beam crossing to reject cosmic rays.

\section{\boldmath{$\lambda_{ ijk}$} Analysis}
\label{par:lambda_anal}

Table~\ref{tab:topologies} shows the possible topologies arising
when the $\mathrm \lambda_{ijk}$ couplings dominate. The different
selections can be classified into four categories as follows: 
$2 \ell + \Emiss$, $4 \ell + \Emiss$, $6 \ell$,
$(\ge 4) \, \ell$ plus possible jets and $\Emiss$. $\Emiss$ (missing energy)
indicates final state neutrinos escaping detection.
After a common preselection,
a dedicated selection is developed for each group, taking 
into account lepton flavours, particle boosts and 
virtual W and Z decay products. 

\begin{table*} [htbp] 
  \begin{center}
  \begin{tabular}{|l|l|l|} \hline 

     \multicolumn{1}{|l} {Direct decays}
    &\multicolumn{1}{l|} { }
    & Selections \\ \hline

     \multicolumn{1}{|l}   {\epem\ra~$\chim\chin \ra$} \rpvpha
    &\multicolumn{1}{l|}  {\hspace{-0.3 cm}{\em $\ell\ell\ell\ell$}$\nu\nu$ } 
    &   4 $\mathrm{\ell}$
      + $\Emiss$                  \\ \hline

   \multicolumn{1}{|l}  { \epem\ra~$\chap\cham \ra$} \rpvpha
   &\multicolumn{1}{l|} {\em\hspace{-0.3 cm}$\ell\ell\ell\ell\ell\ell$}
   &   6   $\mathrm{\ell}$   \\

   \multicolumn{1}{|l}   {} 
   &\multicolumn{1}{l|} {{\em\hspace{-0.3 cm}$\ell\ell\ell\ell$}$\nu\nu$} 
   &   4  $\mathrm{\ell}$ + $ \Emiss$ \\

   \multicolumn{1}{|l}   {} 
   &\multicolumn{1}{l|} {{\em\hspace{-0.3 cm}$\ell\ell$}$\nu\nu\nu\nu$ } 
   &   2  $\mathrm{\ell}$ + $ \Emiss$  \\ \hline 

    \multicolumn{1}{|l}   { \epem\ra~$\slepp\slepm \;\ra$} \rpvpha
   &\multicolumn{1}{l|} {\hspace{-0.3 cm}$\ell\nu\ell\nu$}
   &   2  $\mathrm{\ell}$ + $ \Emiss$  \\ \hline 

    \multicolumn{1}{|l}   { \epem\ra~$\snu\snu \hspace{0.5 cm}\ra$} \rpvpha
   &\multicolumn{1}{l|} {\hspace{-0.3 cm}$\ell\ell\ell\ell$}
   &   4  $\mathrm{\ell}$  + $ \Emiss$ \\ \hline 

     \multicolumn{1}{|l} {Indirect decays}
    &\multicolumn{1}{l|} { }
    &  \\ \hline

     \multicolumn{1}{|l}  { \epem\ra~$\chim\chin_{(n\geq 2)}$} \rpvpha
    &\multicolumn{1}{l|} {\hspace{-0.4 cm} \ra $\;$ cascades }
    & $(\ge 4) \, \mathrm{\ell}$ + (jets) + $ \Emiss$  \\ \hline


     \multicolumn{1}{|l} {\epem\ra~$\chap\cham \ra$ } \rpvpha
    &\multicolumn{1}{l|}  {\hspace{-0.5 cm}
        $\tilde{\chi}_{1(2)}^0\tilde{\chi}_{1(2)}^0 \Wstar\Wstar$}
    & $(\ge 4) \, \mathrm{\ell}$ + (jets) + $ \Emiss$  \\ \hline

    \multicolumn{1}{|l}   { \epem\ra~$\slepp\slepm \;\ra$} \rpvpha
   &\multicolumn{1}{l|} {\hspace{-0.3 cm}$\ell\ell\ell\ell\ell\ell\nu\nu$}
   & $(\ge 4) \, \mathrm{\ell}$ + (jets) + $ \Emiss$  \\ \hline

    \multicolumn{1}{|l}   { \epem\ra~$\snu\snu \hspace{0.5 cm}\ra$} \rpvpha
   &\multicolumn{1}{l|} {\hspace{-0.3 cm}$\ell\ell\ell\ell\nu\nu\nu\nu$}
   &   4  $\mathrm{\ell}$  + $\Emiss$  \\ \hline

  \end{tabular}
  \caption{Processes considered in the 
    $\mathrm \lambda_{ijk}$ coupling analysis
    and corresponding selections. 
    $\protect\chim\protect\chin$ indicates neutralino pair-production 
    with $m = 1,2 \,$ and $n = 1,..,4$. ``Cascades''  refers
    to all possible final state combinations of Table~\ref{tab:decays}.}
  \label{tab:topologies}
  \end{center}
\end{table*}

Events are preselected by requiring at least two charged leptons 
($N_{e,\mu,\tau}$), to reject $\mathrm \qqbar$ events and
hadronic $\mathrm \WpWm$ and ZZ decays. 
Events have to contain at least 3 charged tracks  ($N_{tracks}$) and
4 calorimetric clusters  ($N_{clusters}$) in order to remove 
$\epem \ra \epem, \mpmm$ and purely leptonic $\mathrm \tautau$ 
and $\mathrm \WpWm$ decays. 
The visible energy ($E_{vis}$) is required to be 
less than $90\%$ of $\sqrt s$, in order to reject 
background from lepton pair-production.
If the number of tracks is at least 5,
the cut on the visible energy is not applied, otherwise it
would suppress signal events with six charged leptons.
Back-to-back events, in particular $\mathrm \tautau$, are reduced 
by requiring ${y_{34}}$ to be greater than 0.0002.
For low multiplicity signal events belonging to the
$2 \ell + \Emiss$ topologies,
the following cuts are applied:
events must have at least 2 and less than 6 tracks, between 2 and 15
calorimetric clusters, the visible energy has to be
less than $60\%$ of $\sqrt s$ and $N_{jets8}$ has to be at least 2.
In this case, the preselection
requirement on ${y_{34}}$ is not applied. In addition, for the
$2 \ell + \Emiss$ topologies, the acollinearity and acoplanarity
angles are required to be below $\mathrm 176^\circ$ in order
to reject $\ell^+\ell^-$ background.

Untagged two-photon interactions are removed by requiring
the visible energy to be greater than $20\%$ of $\rts$. 
The polar angle ($\theta_{miss}$) of the missing momentum vector 
has to be between $\mathrm 15^\circ$ and $\mathrm 165^\circ$. 
Background from two-photon interactions is further reduced 
by requiring the missing transverse momentum ($p^T_{miss}$)
to be greater than 7 \GeV.
Tagged two-photon interactions are rejected by requiring the sum
of the energies measured in the small angle calorimeters 
between $\mathrm 1.5^\circ$ and $\mathrm 9.0^\circ$ 
to be less than 10 \GeV.

\begin{figure}[htbp]
  \begin{center}
\hspace{-2.0 cm}
    \includegraphics[width=17.5 cm, height=1.2 cm]{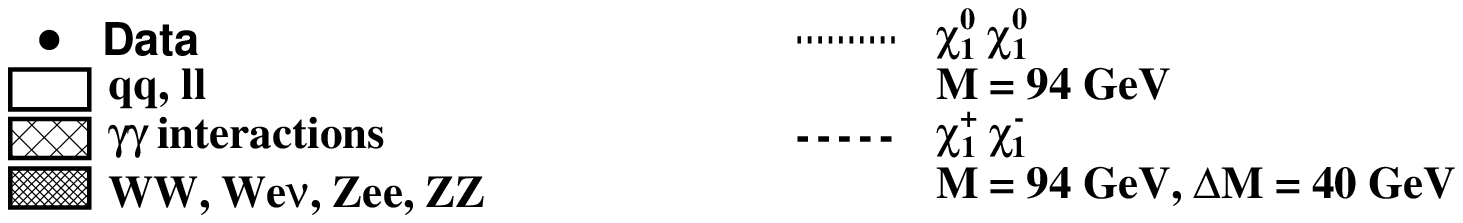}
\end{center}
\begin{center}
    \includegraphics[width=16.5cm, height=16.5cm]{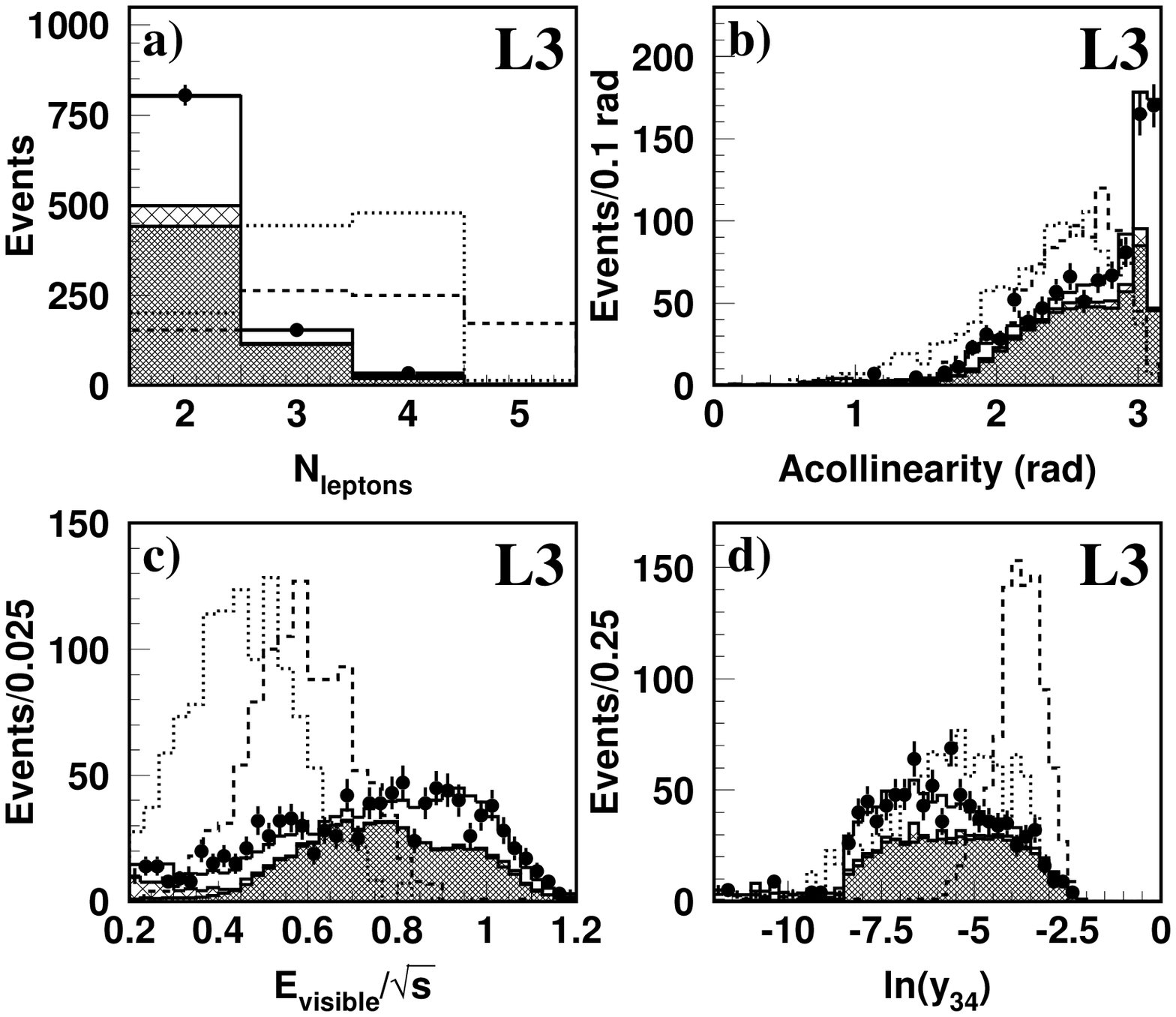}
  \end{center}
  \caption{Data and Monte Carlo
 distributions of a) the number of leptons, b) acollinearity,
 c) the normalised visible
 energy and  d) ln(${y_{34}}$) after the 
 $\mathrm \lambda_{ijk}$ preselection.
 The solid histograms show the expectations for Standard Model 
 processes at $ \rts =$ 189 \GeV.
The dotted and dashed histograms
 show two examples of signal, with dominant coupling $\lambda_{133}$.
The dotted histograms represent the process $\mathrm \epem \ra 
 \protect\chio\protect\chio$, for $\protect\mchi =94$ \GeV,
corresponding to five hundred times the luminosity of the data.
The dashed ones
represent $\mathrm \epem \ra \protect\chap\protect\cham$, with $\protect\mcha = 94$ \GeV{} and
 $\DM = \protect\mcha - \protect\mchi = 40$ \GeV{},
corresponding to twenty times the luminosity.}
  \label{fig:ps_lambda}
\end{figure}

After the preselection is applied, 
995 events are selected in the data sample and $984 \pm 6$ events
are expected from Standard Model processes, of which 398 are
from $\mathrm \WpWm$, 136 from $\mathrm{W^\pm\, e^\mp \nu}$ decays
and 258 from $\qqbar$ events.
Figure \ref{fig:ps_lambda} shows the
number of leptons, acollinearity, normalised visible
energy and ln(${y_{34}}$) distributions after the
preselection.
The data are in good agreement with the Monte Carlo expectations.

The four groups of final selections are shown in 
Tables~\ref{tab:lambda_sel1} to~\ref{tab:lambda_sel3b} of
Appendix~\ref{par:appendix_sel}.
The efficiencies are summarized in Tables~\ref{tab:efficiencies1} and
~\ref{tab:efficiencies2}.
Here and in the following sections we discuss only the results obtained 
for those choices of the generation indices which give the
lowest selection efficiencies. In this way, the quoted results
will be conservatively valid for any ${ijk}$ combination.
In the case of direct R-parity violating decays, the efficiencies 
are estimated for different mass values of the pair-produced
supersymmetric particles. In the case of indirect decays,
the efficiencies are estimated for different masses and $\DM$ ranges.
$\DM$ is defined as the mass difference $M_{susy} - \mchi$, where
$M_{susy}$ can assume different values according to the type of
supersymmetric particle whose decay is considered.

In the case of direct neutralino or chargino decays,
the lowest efficiencies are found for $\lambda_{ijk} = \lamtre$,
due to the presence of taus in the final state.
The efficiencies increase with the neutralino or chargino mass. 
At high masses (greater than 50 \GeV{}), 
six fermions are expected to be isotropically produced 
and can be disentangled from W pair-production background events. 
For low masses, the signal signatures look like
back-to-back jet events and the selection efficiencies are smaller due 
to cuts to reduce the dominant background coming from the two-fermion processes. 
In addition, the efficiencies obtained for low masses
are higher for charginos than for neutralinos 
due to the contribution of the six charged lepton final state.

For indirect chargino decays and for a chargino mass of 94 \GeV,
the efficiencies decrease with increasing $\DM$. 
At high $\DM$, the signal signatures are
very similar to those of W pair-production.
For $\mchim + \mchin = $ 188 \GeV,
the efficiencies of the process $\epem \ra \mathrm \chim \chin \;$ 
($m = 1,2$, $\, n = 2,3,4$) decrease slightly with increasing $\DM$.

In the case of pair-production of scalar charged leptons, followed
by direct decays via $\lambda_{ijk}$, the final state contains 
two leptons plus missing energy. The lepton flavours are given
by the indices $i$ and $j$, independently of the value of $k$,
as shown in Table~\ref{tab:decays}.
The lowest selection efficiency is found for $\lambda_{ijk} = \lambda_{12k}$,
{\it i.e.} for events with electrons and/or muons in the final state.
This is due to the fact that 
it is necessary, in order to reject a potential large background from
lepton pair-production, to select events with
at least 3 calorimetric clusters and with visible
energy below 55\% of $\rts$ (Table~\ref{tab:lambda_sel2}). For this reason,
events with lower multiplicity coming from $\lambda_{12k}$ mediated decays are 
selected with lower efficiency. The efficiency increases with increasing
scalar lepton mass.
 
Direct decays of scalar neutrinos provide four leptons in the final state. 
In this case we have used the
$4 \ell + \Emiss$ and $(\ge 4) \, \ell + $  (jets) $ + \Emiss$
selections, without developing a dedicated one, since these
selections provide, also for $4 \ell$ events,  a good analysis sensitivity
comparable to that of the dedicated selections
for scalar electrons, muons and taus.
Events with scalar neutrino decays into electrons 
and muons are thus selected with lower efficiency than events with decays
into taus. In particular, the smallest efficiency is obtained for
$\lambda_{121}$, which can give rise to the decays 
$\mathrm{\snu_e \ra \mu^- e^+\,}$ and  $\mathrm{\snu_\mu \ra e^- e^+}$. 
This effect is
due to the selection requirements on the energy in the electromagnetic 
calorimeter, which are optimized for selecting events with
$4 \ell + \Emiss$.

Indirect decays of charged scalar leptons ($\slepp\slepm \ra 
\ell^+ \ell^- \chio\chio$)  
provide six leptons plus missing energy in the final state.
The events are selected by means of the 
$(\ge 4) \, \ell + $  (jets) $ + \Emiss$ selections. 
Table~\ref{tab:efficiencies2} shows the efficiencies for scalar
electron, muon and tau pair-production, for different values of 
$\DM = M_{\slepr} - \mchi$.
For large values of $\DM$ the selection efficiencies are smaller
for $\serp\serm$ than for the other channels because of the cut on
the energy in the electromagnetic calorimeter.

Indirect decays of scalar neutrinos ($\snu\snu \ra \nu\nu \chio\chio$)
are selected by the $4 \ell + \Emiss$ selections. The efficiency
decreases with increasing $\DM$, when the two additional neutrinos
can carry a relevant fraction of the total energy.

\vspace{1 cm}
\begin{table*} [htbp]
  \begin{center}
  \begin{tabular}{|l|l|c|c|c|} \hline
     \multicolumn{2}{|c|} {Direct decays }
      &\multicolumn{3}{c|} {Mass values } \\ \hline
     Coupling   & Process  & $M =$ 5--20 \GeV & $M =$ 25--50 \GeV & $M $=55--94
\GeV \\ \hline
 $ \lambda_{133}$   & $\chim\chin$ \rpvpha            &  4\%--13\% & 21\%--35\% & 41\%--51\% \\ \hline
 $ \lambda_{133} $  & $\chap\cham$ \rpvpha             &  8\%--10\% & 15\%--36\% & 43\%--45\% \\ \hline
 $ \lambda'_{311}$  & $\chim\chin$, $\chap\cham$ \rpvpha & 6\%--17\% & 19\%--28\% & 21\%--26\% \\ \hline
 $ \lambda''_{212}$ & $\chim\chin$, $\chap\cham$ \rpvpha & -- & 29\%--38\% & 35\%--54\% \\ \hline

     \multicolumn{2}{|c|} {Indirect decays }
      &\multicolumn{3}{c|} {$\DM$ values } \\ \hline
    Coupling & Process & ${\DM}=$ 5--20 \GeV & ${\DM}=$ 25--50 \GeV & 
               ${\DM}=$ 55--80 \GeV \\ \hline
     $ \lambda_{133}$   & $\chim\chin_{(n\geq 2)}$  \rpvpha
                                        & 50\%--51\%  & 48\%--50\% & 42\%--46\% \\ \hline
$ \lambda_{133}$   & $\chap\cham$ \rpvpha & 47\%--59\%  & 33\%--43\% & 14\%--27\% \\ \hline
$ \lambda'_{311}$  & $\chap\cham$ \rpvpha & 25\%--28\%  & 29\%--30\% & 17\%--23\% \\ \hline
$ \lambda''_{212}$ & $\chim\chin_{(n\geq 2)}$   \rpvpha
                                          & 55\%--59\%  & 59\%--60\% & 47\%--53\% \\ \hline
$ \lambda''_{212}$ & $\chap\cham$ \rpvpha & 54\%--62\%  & 60\%--66\% & 43\%--56\% \\ \hline

  \end{tabular}
  \caption{Efficiency ranges for neutralino and chargino production.
    $\protect\chim\protect\chin$ indicates neutralino pair-production 
    with $m = 1,2 \,$ and $n = 1,..,4$. The efficiencies correspond to
    $ \protect\mchim + \protect\mchin = $ 188 \GeV. For
    direct decays the lowest mass values considered are $\protect\mchi = 5$ \gev{} 
    and $\protect\mcha = 15$ \GeV{}
    for $\lamtre$, $M =$ 15 \gev{} for $ \lambda'_{311}$ and  
    $\protect\mchi = 20$ \gev{} and $\protect\mcha =$ 45 \GeV{} for $\lamuno$. 
    For direct neutralino decays 
    we quote the $\protect\chio\protect\chio$ efficiencies. 
    In the case of indirect decays 
    the chargino selection efficiencies correspond to $\protect\mcha =$ 94 \GeV.}
      \label{tab:efficiencies1}
  \end{center}
\end{table*}

\begin{table*} [htbp]
  \begin{center}
  \begin{tabular}{|l|l|c|c|c|} \hline
     \multicolumn{2}{|c|} {Direct decays }
      &\multicolumn{3}{c|} {Mass values } \\ \hline
     Coupling   & Process  & $M =$ 5--20 \GeV & $M =$ 25--50 \GeV & $M $=55--94
\GeV \\ \hline
    $ \lambda_{12k} $   & $\slepp\slepm$ \rpvpha & --  & -- &  8\%--18\%  \\ \hline
    $ \lambda_{121}$    & $\snu\snu$  \rpvpha     & --  & -- &  6\%--8\%  \\ \hline
     $ \lambda'_{311}$   & $\serp\serm\;$  *  \rpvpha   & -- & 14\%--16\% & 18\%--22\% \\ \hline
     $ \lambda''_{212}$   & $\serp\serm \;$ *  \rpvpha & -- & 32\%--38\% & 35\%--54\% \\ \hline
     $ \lambda''_{212}$   & $\smurp\smurm$ *   \rpvpha & -- & 32\%--38\% & 35\%--54\% \\ \hline
     $ \lambda''_{212}$   & $\staurp\staurm\,$ * \rpvpha & -- & 32\%--38\% & 15\%--44\% \\ \hline

     \multicolumn{2}{|c|} {Indirect decays }
      &\multicolumn{3}{c|} {$\DM$ values } \\ \hline
    Coupling & Process & ${\DM}=$ 5--20 \GeV & ${\DM}=$ 25--50 \GeV & 
               ${\DM}=$ 55--80 \GeV \\ \hline

     $ \lambda_{133}$   & $\serp\serm$   \rpvpha   & 74\%--74\% & 50\%--63\% & 26\%--38\% \\ \hline
     $ \lambda_{133}$   & $\smurp\smurm$  \rpvpha  & 82\%--84\% & 76\%--84\% & 65\%--72\%  \\ \hline
     $ \lambda_{133}$   & $\staurp\staurm$  \rpvpha& 66\%--72\% & 67\%--73\% & 57\%--65\%  \\ \hline
     $ \lambda_{133}$   & $\snu\snu$    \rpvpha  &  52\%--52\% & 39\%--49\%  & 28\%--37\% \\ \hline
     $ \lambda'_{311}$   & $\serp\serm$   \rpvpha   & 30\%--60\% & 50\%--60\% &
  60\%--76\% \\ \hline
     $ \lambda''_{212}$   & $\serp\serm$  \rpvpha    & 37\%--63\% & 66\%--69\% &
  45\%--62\% \\ \hline
     $ \lambda''_{212}$   & $\smurp\smurm$   \rpvpha & 29\%--51\% & 54\%--57\% &
  34\%--50\% \\ \hline
     $ \lambda''_{212}$   & $\staurp\staurm$ \rpvpha & 56\%--66\% & 39\%--62\% &
  15\%--16\% \\ \hline

  \end{tabular}
  \caption{Efficiency ranges for scalar lepton production.
    In the case of direct decays the lowest mass value considered 
    is $M_{\slepr} = 54$ \GeV{}
    for $\lambda_{12k}$, $M_{\snu} = 54$ \GeV{}  for $\lambda_{121}$,
    $M_{\ser} = 45$ \GeV{} for $\lambda'_{311}$ and
    $M_{\slepr} = 30$ \GeV{} for $\lambda''_{212}$.
    For the processes marked with~* we refer to four-body decays, as 
    described in Sections~\ref{par:lambdapri_anal} 
    and~\ref{par:lambdasec_anal}.
    For indirect decays 
    the scalar lepton selection efficiencies correspond to 
    $M_{\slepr} \,(M_{\snu}) \, =$ 94 \GeV.}
      \label{tab:efficiencies2}
  \end{center}
\end{table*}

\newpage
\section{\boldmath{$\lambda'_{ ijk}$} Analysis}
\label{par:lambdapri_anal}

Table~\ref{tab:topologies_lqd} shows the topologies arising
when the $\mathrm \lambda'_{ijk}$ couplings dominate, as well as the
different final states, classified into five categories.
The quark flavour composition is given by the 
{\it j} and {\it k} generation indices of the dominant  
$\lambda'_{ijk}$ coupling.
After a common preselection, a dedicated selection is developed 
for each group, taking into account lepton flavours and charge, 
particle boosts and virtual Z and W decay products. 

\begin{table*} [htbp] 
  \begin{center}
  \begin{tabular}{|l|l|l|} \hline 

     \multicolumn{1}{|l} {Direct decays}
    &\multicolumn{1}{l|} { }
    & Selections \\ \hline

     \multicolumn{1}{|l}   {\epem\ra~$\chim\chin \;\ra$} \rpvpha
    &\multicolumn{1}{l|}  {\hspace{-0.3 cm}{\rm qqqq $\ell\ell$}}
    &    4 jets + 2 $\ell$    \\
     \multicolumn{1}{|l}   { } 
    &\multicolumn{1}{l|}  {\hspace{-0.3 cm}{\rm qqqq $\ell\nu$}}
    &    4 jets + $\ell$ + $\Emiss$     \\ 
     \multicolumn{1}{|l}   { } 
    &\multicolumn{1}{l|}  {\hspace{-0.3 cm}{\rm qqqq $\nu\nu$}}
    &    4 jets + $\Emiss$     \\ \hline

     \multicolumn{1}{|l}   {\epem\ra~$\chap\cham \;\ra$} \rpvpha
    &\multicolumn{1}{l|}  {\hspace{-0.3 cm}{\rm qqqq $\,\ell\ell$}}
    &    4 jets + 2 $\ell$    \\
     \multicolumn{1}{|l}   { } 
    &\multicolumn{1}{l|}  {\hspace{-0.3 cm}{\rm qqqq $\,\ell\nu$}}
    &    4 jets + $\ell$ + $\Emiss$     \\ 
     \multicolumn{1}{|l}   { } 
    &\multicolumn{1}{l|}  {\hspace{-0.3 cm}{\rm qqqq $\,\nu\nu$}}
    &    4 jets + $\Emiss$     \\ \hline

     \multicolumn{1}{|l} {Indirect decays}
    &\multicolumn{1}{l|} { }
    &  \\ \hline

     \multicolumn{1}{|l}  { \epem\ra~$\chim\chin_{(n\geq 2)}$} \rpvpha
    &\multicolumn{1}{l|}  {\hspace{-0.4 cm} \ra $\;$ cascades }
    & multijets (+ leptons) (+ $\Emiss$) \\ \hline 

     \multicolumn{1}{|l} {\epem\ra~$\chap\cham \ra$ } \rpvpha
    &\multicolumn{1}{l|}  {\hspace{-0.5 cm}
        $\tilde{\chi}_{1(2)}^0\tilde{\chi}_{1(2)}^0 \Wstar\Wstar$}
    & multijets (+ leptons) (+ $\Emiss$) \\ \hline 

    \multicolumn{1}{|l}   { \epem\ra~$\serp\serm \;\ra$}
   &\multicolumn{1}{l|} {\hspace{-0.3 cm}$\chio\chio$ ee  } \rpvpha
   &   4 jets + 2--4 $\mathrm{\ell}$ (+ $\Emiss$)  \\ \hline

  \end{tabular}
  \caption{Processes considered in the 
    $\mathrm \lambda'_{ijk}$ coupling analysis and corresponding
    selections. For masses below 50~\gev{}
    or small $\DM$ values not all jets in the event can be resolved.
    $\protect\chim\protect\chin$ indicates neutralino pair-production 
    with $m = 1,2 \,$ and $n = 1,..,4$. ``Cascades''  refers
    to all possible final state combinations of Table~\ref{tab:decays}.}
  \label{tab:topologies_lqd}
  \end{center}
\end{table*}

Events are preselected by requiring at least 4 charged tracks
and 5 calorimetric clusters in order to remove 
$\epem \ra \epem, \mpmm$ and purely leptonic $\mathrm \tautau$ 
and $\mathrm \WpWm$ decays. 
For centre-of-mass energies 
above the Z peak a large fraction of background events contains a
hard initial state radiation (ISR) photon. In order to remove
these events the polar angle of the missing momentum has to be
between $\mathrm 7^\circ$ and $\mathrm 173^\circ$.

Untagged two-photon interactions are removed by requiring
the energy in a cone of $\mathrm 12^\circ$ half-opening angle around the
beam axis  not to exceed 20\% of the total visible energy. 
In addition, the visible energy must be greater than $30\%$ of 
$\rts$.
Tagged two-photon interactions are rejected as explained in
Section~\ref{par:lambda_anal}.

In this analysis no attempt is made to identify quark flavours.
However, the efficiency is found to be slightly higher for
events containing b quarks than for events with light quarks.
When  $i = 3$, the decay products
of the neutralino pair contain $\tautau$, $\tau \nu_\tau$
or $ \nu_\tau \bar\nu_\tau$. Since taus are selected with
lower efficiency with respect to electrons or muons,
this choice of the first generation index provides a conservative
estimate for the signal efficiency.
The lowest selection efficiency is thus found for
the choice $\lambda'_{ijk} = \lambda'_{311}$,
that will be quoted in the following. 

\begin{figure}[htbp]
  \begin{center}
\hspace{-2.0 cm}
    \includegraphics[width=17.5 cm, height=1.2 cm]{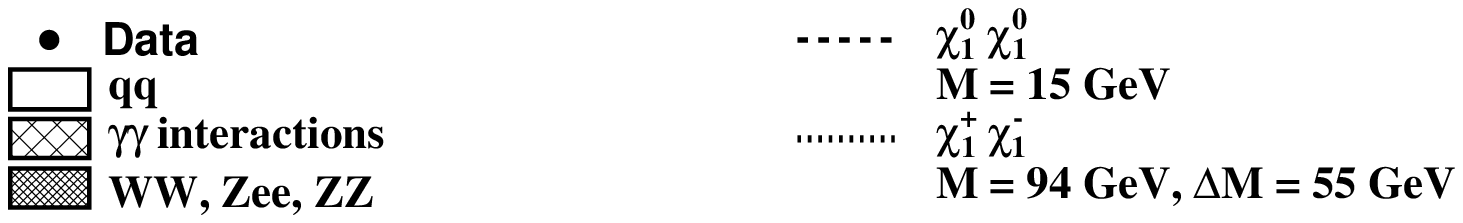}
\end{center}
\begin{center}
    \includegraphics[width=16.5cm, height=16.5cm]{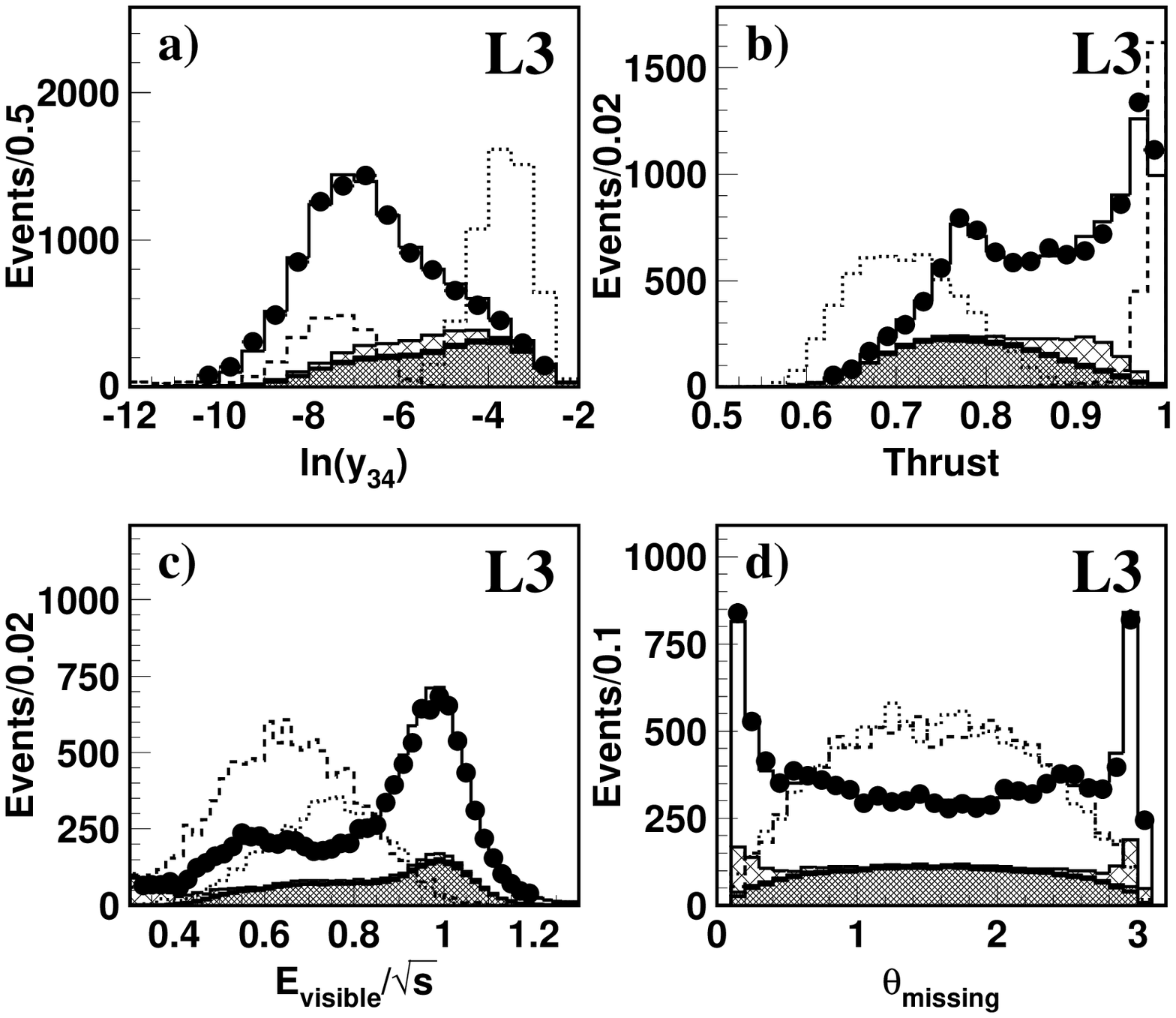}
  \end{center}
  \caption{Data and Monte Carlo
distributions of a) ln(${y_{34}}$), b) thrust, c) 
    normalised visible energy 
 and d) polar angle of the missing momentum after the
 $\lambda'_{ijk}$ preselection. 
 The solid histograms show the expectations for Standard Model 
 processes at $ \rts =$ 189 \GeV.
 The dashed and dotted histograms show two examples of 
 signal, with coupling $\lambda'_{311}$.
The dashed histograms represent the process
$\mathrm \epem \ra \protect\chio\protect\chio \ra $ 4 jets $\nu\bar\nu$, 
with $ \protect\mchi = 15$ \GeV{}, corresponding to
thirty times the luminosity of the data.
The dotted ones represent
$\mathrm \epem \ra \protect\chap\protect\cham$, with $\protect\mcha = 94$ \GeV{} and
 $\DM = \protect\mcha - \protect\mchi = 55$~\GeV{}, with subsequent $\protect\chio\protect\chio$ decays
into 4 jets $\nu\bar\nu$, corresponding to one hundred 
times the luminosity.}
  \label{fig:ps_lambdaprime}
\end{figure}

After the preselection is applied, 
11099 events are selected in the data sample and 
$11022 \pm 34$ events are expected from Standard Model processes, 
of which 7677  events are from $\mathrm \qqbar$, 2481 
from $\mathrm \WpWm$ decays and 674 from 
hadronic two-photon interactions.
Figure \ref{fig:ps_lambdaprime} shows the ln(${y_{34}}$),
thrust, normalised visible
energy and polar angle of the missing momentum
distributions after the preselection. 
The data are in good agreement with the Monte Carlo expectations.

The five groups of final selections are shown in 
Tables~\ref{tab:lambdaprime_sel1} to~\ref{tab:lambdaprime_sel7} of
Appendix~\ref{par:appendix_sel}.
The lowest selection efficiencies correspond to
$\lambda'_{ijk} = \lambda'_{311}$, and are summarized in 
Tables~\ref{tab:efficiencies1} and~\ref{tab:efficiencies2}.

In the case of neutralino and chargino direct decays
the lowest selection efficiency is obtained
in the intermediate mass region (for mass values around 50 \gev)
where the $\mathrm \WpWm $ background
can not be efficiently rejected by a cut on the thrust.

For indirect chargino decays and for a chargino mass of 94 \GeV{} 
the efficiencies decrease for high values of $\DM$, since
in this region the signal signatures are
very similar to those of W pair-production.

In the case of dominant $\lambda'_{ijk}$ couplings, 
scalar leptons can decay indirectly: 
$\slepr \ra \ell \, \chio \ra \ell \ell$qq, $\, \ell \nu$qq.
The efficiencies for indirect decays are listed in 
Table~\ref{tab:efficiencies2}. 
Only $\serp\serm$ production is  considered. The efficiency is highest for large $\DM$, when 
the two energetic electrons give a clear signature.

Supersymmetric partners of the right-handed leptons have no direct
two-body decays via $\lambda'_{ijk}$ couplings (Table~\ref{tab:decays}).  
However, when scalar leptons are lighter than $\chio$, 
the four-body~\cite{superp2} decay $\slepr \ra \ell \ell$qq 
or $\slepr \ra \ell \nu$qq can occur.
This leads to the same final states as those resulting from $\slepr$ indirect 
decays, but with virtual $\chio$ production.
Since non-resonant four-body decays are not implemented in the 
generator~\cite{susygen}, we have used the
results of the indirect decay analysis, with a procedure analogous
to the one described at the end of  Section~\ref{par:lambdasec_anal}.

\section{\boldmath{$\lambda''_{ ijk}$} Analysis}
\label{par:lambdasec_anal}

Table~\ref{tab:topologies_udd} shows the topologies arising
when the $\mathrm \lambda''_{ijk}$ couplings dominate.
The flavour composition depends on the generation indices.
In the case of neutralino and chargino pair-production, 
the different topologies can be classified into two groups:
multijets and multijets with leptons and/or missing energy.
After a common preselection, dedicated selections are
developed for each group, depending on the particle boosts,
the $\DM$ values and the 
virtual W decay products.
The following process is
also considered: $\epem\ra \slepp\slepm \ra {\rm qqqqqq} \ell^+\ell^-$,
for which a third group of selections is performed, taking
into account the lepton flavour.

\begin{table*} [htbp] 
  \begin{center}
  \begin{tabular}{|l|l|l|} \hline 

     \multicolumn{1}{|l} {Direct decays}
    &\multicolumn{1}{l|} { }
    & Selections \\ \hline

     \multicolumn{1}{|l}   {\epem\ra~$\chim\chin \;\ra$} \rpvpha
    &\multicolumn{1}{l|}  {\hspace{-0.3 cm}{\rm qqqqqq}}
    &    multijets                  \\ \hline

   \multicolumn{1}{|l}  { \epem\ra~$\chap\cham \ra$} \rpvpha
   &\multicolumn{1}{l|}   {\hspace{-0.3 cm}{\rm qqqqqq}}
   &   multijets  \\   \hline

     \multicolumn{1}{|l} {Indirect decays}
    &\multicolumn{1}{l|} { }
    &  \\ \hline

     \multicolumn{1}{|l}  { \epem\ra~$\chim\chin_{(n\geq 2)} \ra$} \rpvpha 
    &\multicolumn{1}{l|}  {\hspace{-0.3 cm}{\rm qqqqqq qq}}
    & multijets   \\ 

     \multicolumn{1}{|l}  { } 
    &\multicolumn{1}{l|}   {\hspace{-0.3 cm}{\rm qqqqqq $\ell\ell$}}
    & multijets + lepton(s)  \\ 

     \multicolumn{1}{|l}  { } 
    &\multicolumn{1}{l|}   {\hspace{-0.3 cm}{\rm  qqqqqq $\nu\nu$}}
    & multijets  \\ \hline

     \multicolumn{1}{|l} {\epem\ra~$\chap\cham \ra$ } \rpvpha
    &\multicolumn{1}{l|}  {\hspace{-0.5 cm}{qqqqqq  qqqq}}
    & multijets \\

     \multicolumn{1}{|l} { } 
    &\multicolumn{1}{l|}   {\hspace{-0.5 cm}{qqqqqq  qq $\ell\nu$}}
    & multijets + lepton(s) \\

     \multicolumn{1}{|l} { } 
    &\multicolumn{1}{l|}   {\hspace{-0.5 cm}{qqqqqq $\ell\ell\nu\nu$}}
    & multijets + lepton(s)  \\ \hline

    \multicolumn{1}{|l}   { \epem\ra~$\slepp\slepm \;\ra$} \rpvpha
   &\multicolumn{1}{l|} {\hspace{-0.6 cm} qqqqqq $\ell\ell $}
   &   6 jets + 2 $\mathrm{\ell}$   \\ \hline

  \end{tabular}
  \caption{Processes considered in the 
    $\mathrm \lambda''_{ijk}$ coupling analysis and corresponding
    selections. For masses below 50~\gev{}
    or small $\DM$ values not all jets in the event can be resolved.
    $\chim\protect\chin$ indicates neutralino pair-production 
    with $m = 1,2 \,$ and $n = 1,..,4$.
    For final states with neutrinos we use selections with no explicit
    missing energy requirement, because for those topologies $\Emiss$
    is small.}
  \label{tab:topologies_udd}
  \end{center}
\end{table*}

The preselection of the $\mathrm \lambda''_{ijk}$
analysis aims at selecting well balanced hadronic events.
Low multiplicity events, such as leptonic Z and W decays, are
rejected by requiring at least 13 calorimetric clusters.
At least one charged track has to be present.
The visible energy has to be greater than 70\% of $\sqrt s$. 
The energy imbalances, parallel ($E_{par}$) and perpendicular 
($E_{perp}$) to the beam direction, are required to be less 
than 20\% of the visible energy. 
 Unbalanced events
with an ISR photon in the beam pipe are removed by means of
the requirement on the parallel energy imbalance. 
In order to reject events with an ISR photon seen in the
detector, the invariant mass of the hadronic system ($\rtsp$) has to
be greater than 80\% of $\rts$. 
In order to remove background contributions from two-photon interactions,
the energy in a cone of $\mathrm 12^\circ$ half-opening angle around the
beam axis has not to exceed 30\% of the total visible energy. 
Furthermore, the thrust axis is required to be well contained in the
detector with a polar angle ($\theta_T$) between  $\mathrm 8^\circ$
and  $\mathrm 172^\circ$.

\begin{figure}[htbp]
  \begin{center}
\hspace{-2.0 cm}
    \includegraphics[width=17.5 cm, height=1.2 cm]{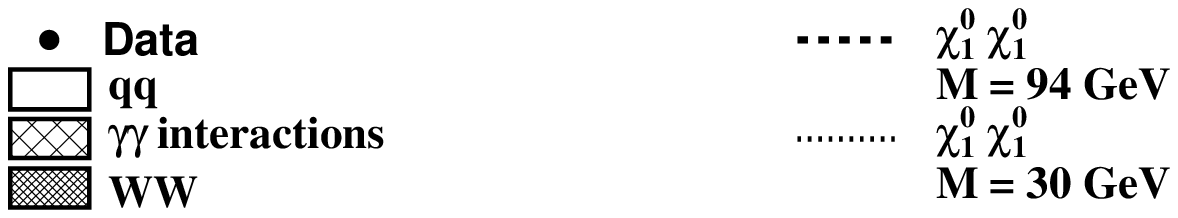}
\end{center}
\begin{center}
    \includegraphics[width=16.5cm, height=16.5cm]{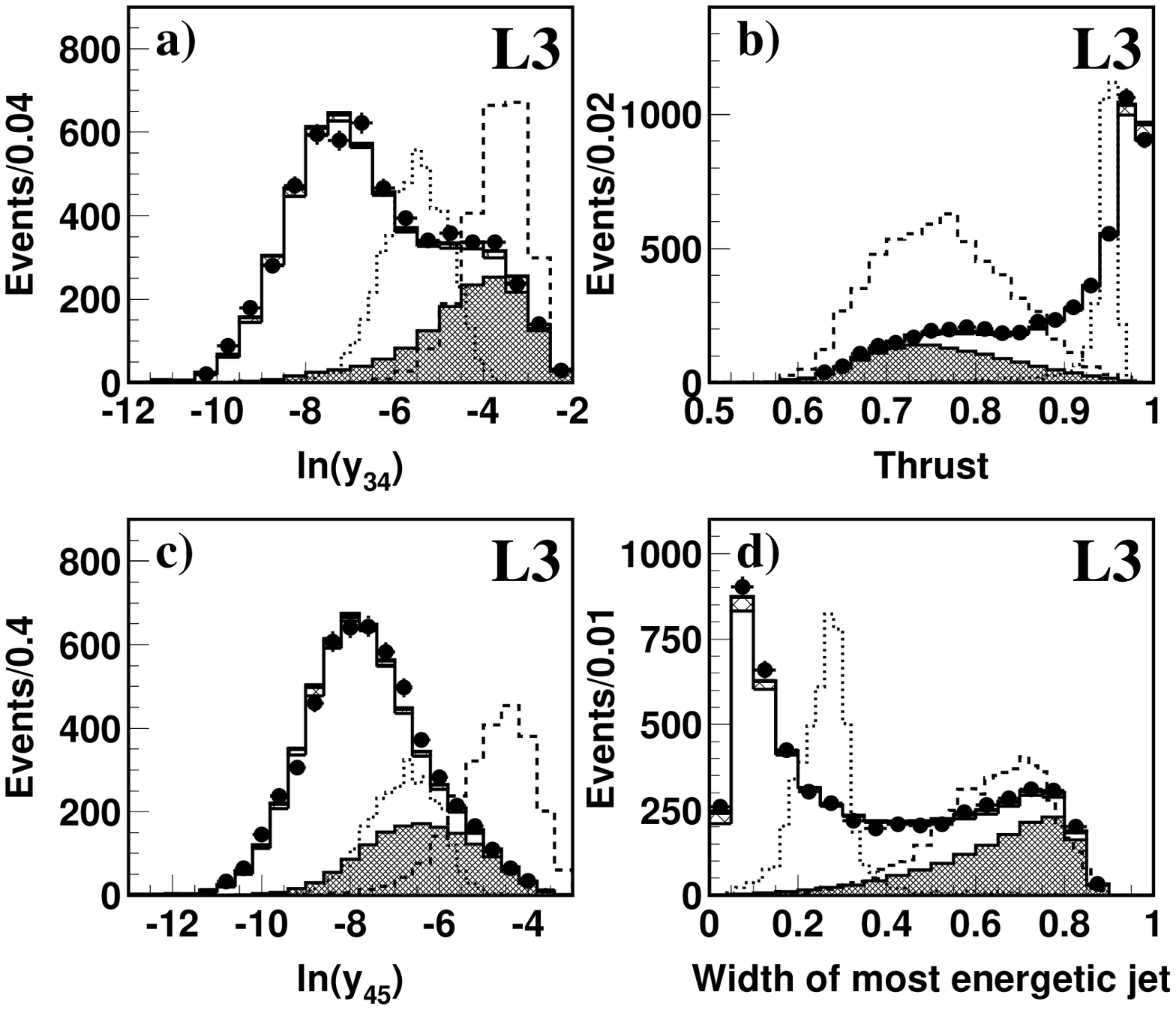}
  \end{center}
  \caption{Data and Monte Carlo
distributions of a) ln(${y_{34}}$), b) thrust, c) ln(${y_{45}}$) 
 and d) width of the most energetic jet after the
 $\lambda''_{ijk}$ preselection. 
 The solid histograms show the expectations for Standard Model 
 processes at $ \rts =$ 189 \GeV.
 The dashed and dotted histograms show two examples of 
 signal, with dominant coupling $ \lamuno$, corresponding to decays into
c, d and s quarks.
The dashed histograms represent the process
 $\mathrm \epem \ra \protect\chio\protect\chio$, with $ \protect\mchi = 94$ \GeV{}, corresponding to
one thousand times the luminosity of the data. 
 The dotted ones represent the same process, with
 $ \protect\mchi = 30$ \GeV{}, corresponding to fifteen times the luminosity.}
  \label{fig:ps_lambdasec}
\end{figure}

Also in this analysis no attempt is made to identify quark flavours.
However, the efficiency is found to be slightly lower for events containing
light quarks than for events with b quarks.
Therefore, only the results obtained by the 
choice $\mathrm \lambda''_{ijk} = \lamuno$ will be quoted
in the next sections. 
After the preselection is applied, 
5492 events are selected in the data sample and $5463 \pm 32$
are expected from Standard Model processes, 
of which 3803  are $\mathrm \qqbar$ and 1431 are
$\mathrm \WpWm$ events.
Figure~\ref{fig:ps_lambdasec} shows the
ln($y_{34}$), thrust, ln($y_{45}$) 
and width of the most energetic jet ($W_{jet1}$)
distributions after the preselection.
The width of a jet is defined as $ {p}_{T}^{jet} / {E^{jet}_{ }}$,
where the event is clustered into exactly two jets,
ordered with decreasing energies,
and ${p}_{T}^{jet}$ is the sum of the projections of the particle 
momenta
on to a plane perpendicular to the jet axis.
There is good agreement between data and Monte Carlo expectations.

The three groups of final selections are shown in 
Tables~\ref{tab:lambdasec_sel1} to~\ref{tab:lambdasec_sel3} of
Appendix~\ref{par:appendix_sel}.
The efficiencies are summarized in Tables~\ref{tab:efficiencies1}
and~\ref{tab:efficiencies2}.

For direct neutralino and chargino decays, the efficiencies increase 
with increasing mass of the supersymmetric particle. 
At high masses, the six quarks are expected to be isotropically produced,
while for low mass values signal events are produced back-to-back
and are selected with lower efficiencies due to cuts required to reduce
the dominant background coming from the two-fermion processes. 

In the case of indirect decays and for a  chargino mass of 94 \GeV{} 
the efficiencies slightly decrease at high values of $\DM$, where
the signal signatures are very similar to those of $\WpWm$ background.
Leptons and neutrinos from 
virtual W decays
can carry a large fraction of the event energy when $\DM$ is large, 
leading to lower selection efficiencies.

For $\mchim + \mchin = $ 188 \GeV,
the efficiencies of the process $\epem \ra \mathrm \chim \chin \;$ 
($m = 1,2$, $\, n = 2,3,4$) decrease slightly with increasing $\DM$.

Scalar leptons can decay indirectly: 
$\slepr \ra \ell \, \chio \ra \ell$qqq.
The efficiencies for these indirect decays are also listed in 
Table~\ref{tab:efficiencies2}. 

For scalar electron and scalar muon decays,
the efficiencies are highest for medium $\DM = M_{\slepr} - \mchi$ 
values, where events have a high multiplicity satisfying the multijet 
selections and two energetic leptons which are well identified. 

Supersymmetric partners of the right-handed leptons have no direct
two-body decays via $\lambda''_{ijk}$ couplings (Table~\ref{tab:decays}).  
However, when scalar leptons are lighter than $\chio$, 
the four-body~\cite{superp2} decay $\slepr \ra \ell$qqq can occur and
this provides the same final state $\ell$qqq as that resulting from indirect
decays, but with virtual $\chio$ production.

The non-resonant four-body decay is not implemented in the 
generator, which allows only scalar lepton decays via
on-shell neutralino production. For this reason, we use the
results of the indirect decay analysis, performing a scan
over all neutralino mass values between 0
and $M_{\slepr}$, for each value of $M_{\slepr}$. 
The resulting lowest efficiency is conservatively quoted 
in the following for 
four-body decays. It is found in most cases for $\mchi \simeq M_{\slepr}$,
that the resulting low energy lepton can not be separated from a nearby
jet.
For scalar taus with masses above 70~\gev{}
the lowest efficiency is found for high $\DM$ values, as in the case
of indirect decays.

\section{Model Independent Results}
\label{par:mod_ind_results}

Table~\ref{tab:tab7} and Table~\ref{tab:tab8} show the number of 
candidates and expected background events for the different
selections and processes, respectively.
The same process may give rise to different final states
(such as chargino direct decays via $\lambda_{ijk}$) or the
same final state (like ``multijets'') can be present as a decay
product of more than one process.

\begin{figure}[htbp]
  \begin{center}
\vspace{1.0 cm}
    \includegraphics[width=\figwidth]{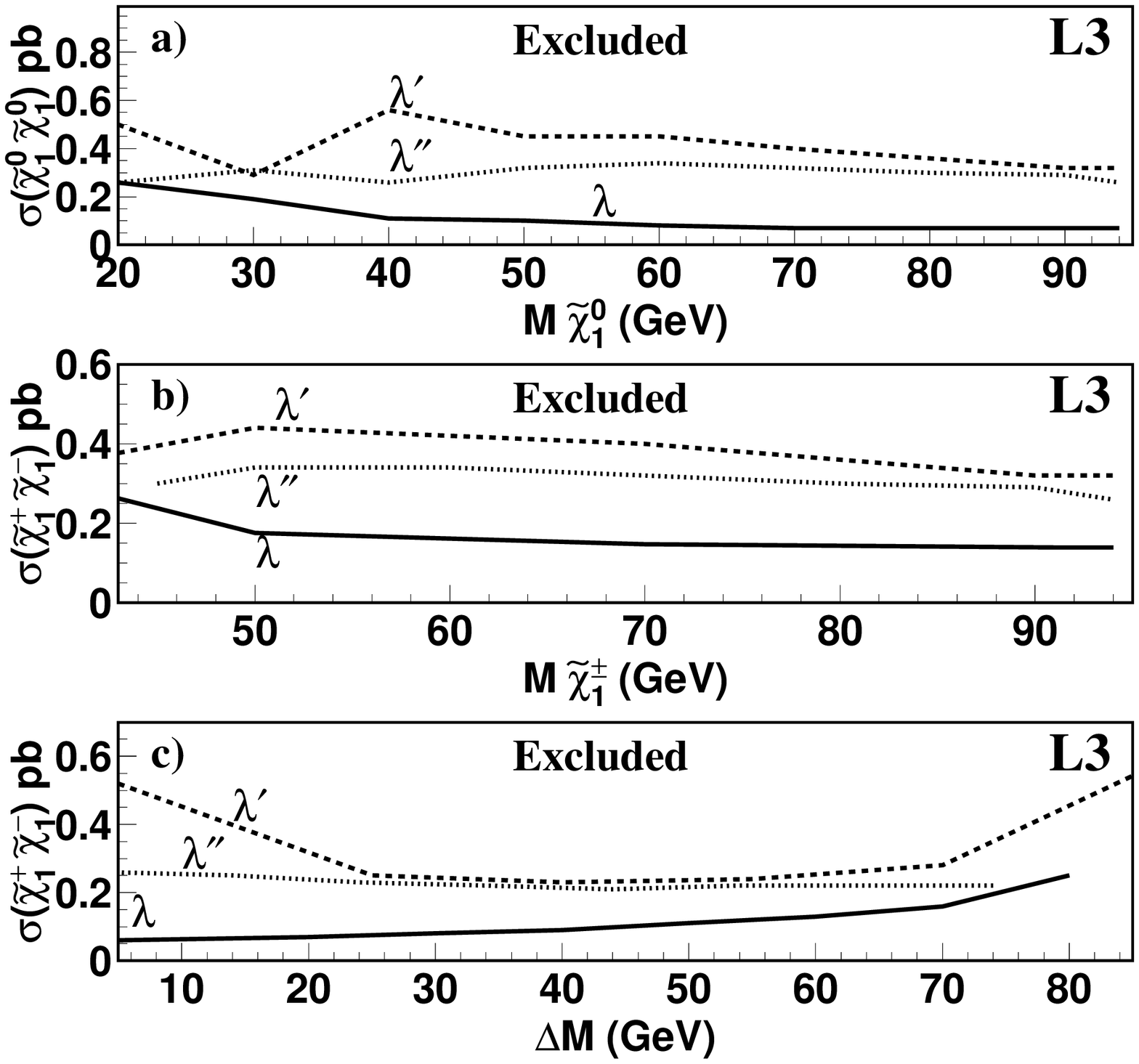}
  \end{center}
  \caption{
     95\% C.L. upper limits on: a) the neutralino pair-production cross
    section as a function of the neutralino mass;
    the chargino pair-production cross
    section b) as a function of the chargino mass, 
    in the direct decay mode and c) 
    as a function
    of $\DM = \protect\mcha - \protect\mchi$, for $\protect\mcha = 94$ \GeV, 
    in the indirect decay mode. 
    The solid lines show the limits
    obtained by the $\lamtre $ analysis, 
    the dotted lines show those obtained by 
    the $\lambda'_{311}$ analysis
    and the dashed lines show those obtained 
    by the $ \lamuno $ analysis. }
  \label{fig:cross_sec1}
\end{figure}

\begin{figure}[htbp]
  \begin{center}
\vspace{4cm}
    \includegraphics[width=\figwidth, height=11cm]{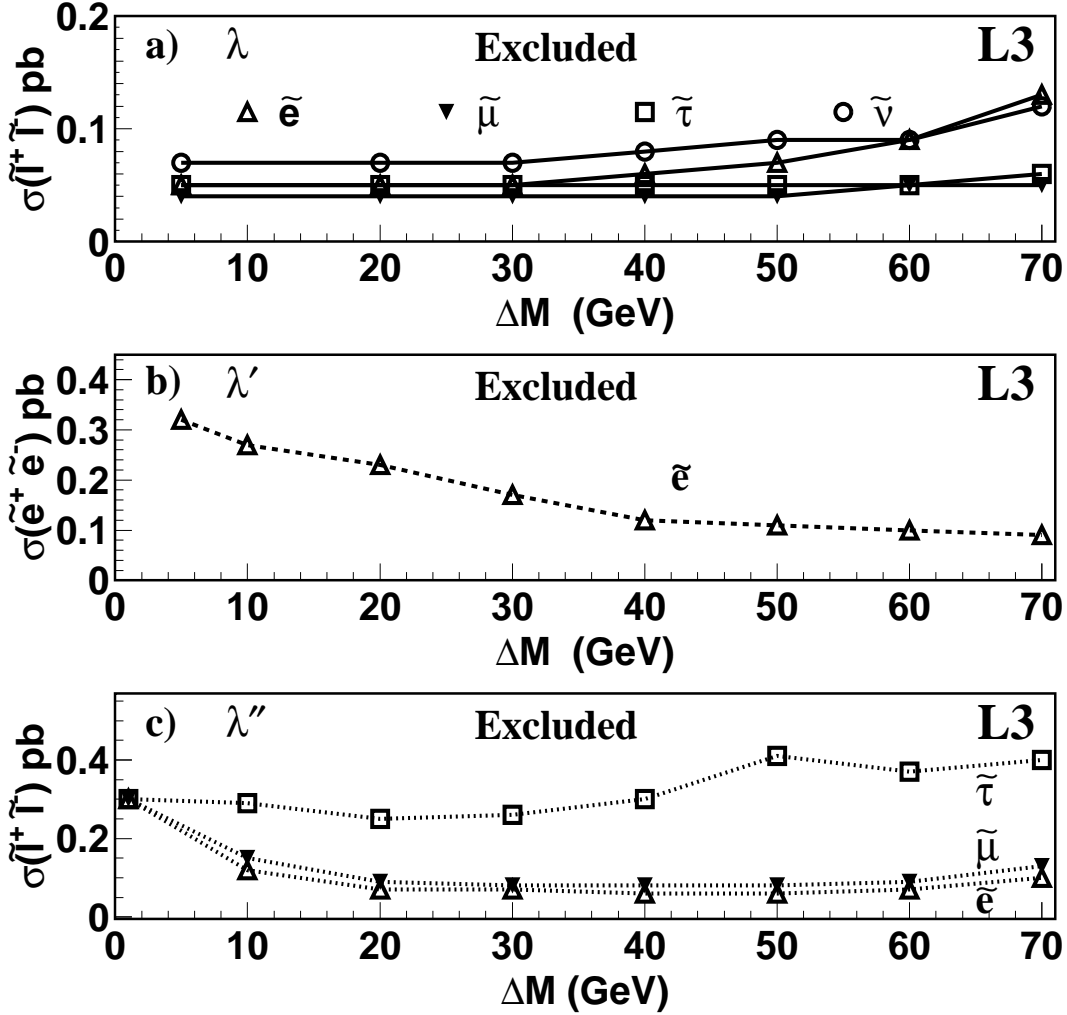}
  \end{center}
  \caption{
     95\% C.L. upper limits on the scalar lepton pair-production cross
    section, in the indirect decay mode, for
    $M_{\slepr} =$ 94 \gev{} and as a function of 
    $\DM = M_{\slepr} - \protect\mchi$,
    for: a) $\lambda_{ijk}$, b) $\lambda'_{ijk}$ and c) $\lambda''_{ijk}$.
    The solid lines show the limits
    obtained by the $\lambda_{133} $ analysis for scalar leptons.
    The dashed line shows the limit obtained by the $\lambda'_{311}$ 
    analysis for scalar electrons.
    The dotted lines show those obtained by 
    the $\lamuno$ analysis for scalar electrons, muons
    and taus.}
  \label{fig:cross_sec3}
\end{figure}

No excess of events is observed. 
Therefore upper limits are set on the neutralino, chargino 
and scalar lepton production cross
sections assuming direct or indirect R-parity violating decays. 
Figure~\ref{fig:cross_sec1} shows the 95\% confidence level
(C.L.) upper limits on neutralino and chargino pair-production 
cross~sections. The 95\% C.L.
upper limits on scalar lepton pair-production cross~sections are shown in
Figures~\ref{fig:cross_sec2} and ~\ref{fig:cross_sec3}.
These limits are derived by taking into account the estimated
background contamination.
Systematic uncertainties on the signal efficiency 
are dominated by Monte Carlo statistics.
The typical relative error is 5\% and it is taken into account in the
calculations of the signal upper limits~\cite{upperlimit}.

\begin{table*} [htbp]
  \begin{center}
  \begin{tabular}{|l|l|c|r|} \hline 

             Coupling &  Selection  & $ N_{back} $ & $ N_{data} $ \\ \hline
     $ \lambda_{ijk}$  & $4\ell + \Emiss$   & 2.0 $\pm$ 0.2     &  3   \\ \cline{2-4}
     & $(\geq 4) \,\ell +$ (jets) $+ \Emiss \,$ & 3.8 $\pm$  0.3   &  2   \\ \cline{2-4}
                       & $2\ell + \Emiss$   & 18  $\pm$  1 \phoescl & 18   \\ \cline{2-4}
                       & $6\ell$            & 0.26  $\pm$  0.05  &  0   \\ \hline

 $   \lambda'_{ijk}$   & 4 jets $ +\, 2 \tau$   & 30.0 $\pm$  0.7\phoescl\phoescl & 26   \\ \cline{2-4}
                       & 4 jets $ + \,\Emiss$     & 28.9 $\pm$  0.8\phoescl\phoescl & 31   \\ \cline{2-4}
                       & 4 jets + $\tau$ + $\Emiss$ & 29.4 $\pm$  1.2\phoescl\phoescl & 25  \\ \cline{2-4} 
                       & Multijets + lepton(s)  & 6.1 $\pm$  0.2 & 8   \\ \cline{2-4}
                       & Multijets + $\Emiss$   & 65 $\pm$  1\phoescl & 68   \\ \cline{2-4}
                       & Multijets + lepton(s) + $\Emiss$ & 11.7 $\pm$  0.2\phoescl\phoescl & 10  \\ \cline{2-4} 
                       & Multijets              & 194  $\pm$  1 \phouno  & 187   \\ \cline{2-4} 
                       & Scalar leptons         & 26.4  $\pm$  0.6\pho & 27  \\ \hline

   $ \lambda''_{ijk}$  & Multijets ($\mchi = $ 20--30 \gev)  & 47 $\pm$  1\phoescl\phoescl    & 42  \\\cline{2-4}
                       & Multijets ($\mchi = $ 30--40 \gev)  & 79 $\pm$  1\phoescl\phoescl     & 81  \\\cline{2-4}
                       & Multijets ($\mchi = $ 40--50 \gev)  & 48.1 $\pm$  0.9 \phoescl & 47  \\\cline{2-4}
                       & Multijets ($\mchi = $ 50--60 \gev)  & 98 $\pm$  1     \phoescl & 100 \\\cline{2-4}
                       & Multijets               & 194 $\pm$  1  \phouno  & 187 \\\cline{2-4}
                       & Multijets + lepton(s) (Semileptonic)    & 1.6 $\pm$  0.2  & 3   \\\cline{2-4}
                       & Multijets + lepton(s) (Leptonic)        & 3.2  $\pm$  0.1  & 3   \\ \cline{2-4}
                       & Scalar leptons          &   154 $\pm$  1 \phouno & 157   \\ \hline

  \end{tabular}
  \caption{Number of observed data ($N_{data}$) and expected background 
  ($N_{back}$) events  for the different selections. 
   A process can give rise
    to several topologies, or the same topology may occur in
    more than one final state.
    The error on the expected background is due to Monte Carlo statistics.}
 \label{tab:tab7}
  \end{center}
%
  \begin{center}
  \begin{tabular}{|l|l|c|r|} \hline 
             Coupling &  Process  & $N_{back}$ & $N_{data}$ \\ \hline
$ \lambda_{ijk}$  & $\chio\chio$ \rpvpha & 2.0 $\pm$ 0.2 &  3   \\ \cline{2-4}
                  & $\chim\chin$ \rpvpha & 5.8 $\pm$ 0.4 &  5   \\ \cline{2-4}
      & $\chap\cham$ (indirect) \rpvpha  & 3.8 $\pm$ 0.3 &  2   \\ \cline{2-4}
        & $\chap\cham$ (direct) \rpvpha  & 20  $\pm$ 1\phoescl  & 21   \\ \cline{2-4}
       & $\slepp\slepm$ (direct) \rpvpha & 18  $\pm$ 1\phoescl  & 18   \\ \cline{2-4}
                  & $\snu\snu$  \rpvpha  & 5.8 $\pm$ 0.4 &  5   \\ \hline
$   \lambda'_{ijk}$ & $\chio\chio$ \rpvpha &    77 $\pm$ 2\phoescl & 70   \\ \cline{2-4}
                    & $\chap\cham$ \rpvpha &   262 $\pm$ 2 \phoescl\phoescl& 257  \\ \cline{2-4}
                    & $\serp\serm$ \rpvpha &  26.4 $\pm$ 0.6\phoescl & 27   \\ \hline
$  \lambda''_{ijk}$ & $\chio\chio$ \rpvpha &   357 $\pm$ 3\phouno\phoescl & 353  \\ \cline{2-4}
                    & $\chap\cham$ \rpvpha &   197 $\pm$ 1\phouno\phoescl & 193   \\ \cline{2-4}
                    & $\slepp\slepm$\rpvpha&   154 $\pm$ 1\phouno\phoescl & 157   \\ \hline
  \end{tabular}
  \caption{Number of observed data ($N_{data}$) and expected background 
  ($N_{back}$) events  for the different processes. 
    Details on the selection of each topology are given in 
    Table~\ref{tab:tab7}. The error on the
     expected background is due to Monte Carlo statistics.}
 \label{tab:tab8}
  \end{center}
\end{table*}

\section{Interpretation in the MSSM}
\label{par:MSSM}

The results are also interpreted as excluded 
regions in the MSSM parameter space. In
the MSSM framework, neutralino and chargino masses, couplings and
cross sections depend on the gaugino mass parameter, ${M_2}$, 
the higgsino mass mixing parameter, $\mu $, the ratio
of the vacuum expectation values of the two Higgs doublets, $\tan\beta$, and
the common mass of the scalar particles at the GUT scale, ${m_0}$.
Therefore the excluded regions in the (${M_2, \mu}$) plane 
are a function of the values of ${m_0}$ and $\tan\beta$.
The results presented in this section are obtained performing a scan over 
 $\,0 \leq M_2 \leq 1000$ \GeV,
$- 500$ \GeV{} $\leq \mu \leq$ 500 \GeV, $0 \leq m_0 \leq 500$ \gev{}
and $0.7 \leq \tan\beta \leq 40$.
They do not depend on the value of the
trilinear coupling in the Higgs sector, $A$.

A point in the MSSM parameter space is excluded if the total number
of expected events is greater than the combined upper limit at 95\% C.L.
on the number of signal events. Neutralino, chargino and scalar
lepton analyses are combined since several processes can occur at a given
point.

In addition to the limits obtained with this analysis, we take into account
the constraints from the L3 cross section measurements at the Z pole.
A point in the MSSM parameter space is excluded at 95\% C.L.
by Z lineshape measurements if:
\begin{equation}{(\frac{\sigma_{SUSY}}{\sigma_{\mathrm Z}}}) \;  
     {\Gamma_{\mathrm Z}} > {\Gamma_{LIM}}\,,
\end{equation}
where $\sigma_{SUSY}$ is the sum of the
pair-production cross sections  
of supersymmetric particles at the Z pole, 
calculated with {\tt SUSYGEN}.  
$\sigma_{\mathrm Z}$ is the measured total Z cross section,
$\Gamma_{\mathrm Z}$ and 
$\Gamma_{LIM} = 22\MeV$ are 
the measured total Z width and 
the 95\% C.L. upper limit on possible 
non-Standard Model contributions to the total Z width~\cite{gammalim}. 

\begin{figure}[htbp]
  \begin{center}
    \includegraphics[width=8 cm, height=8cm]{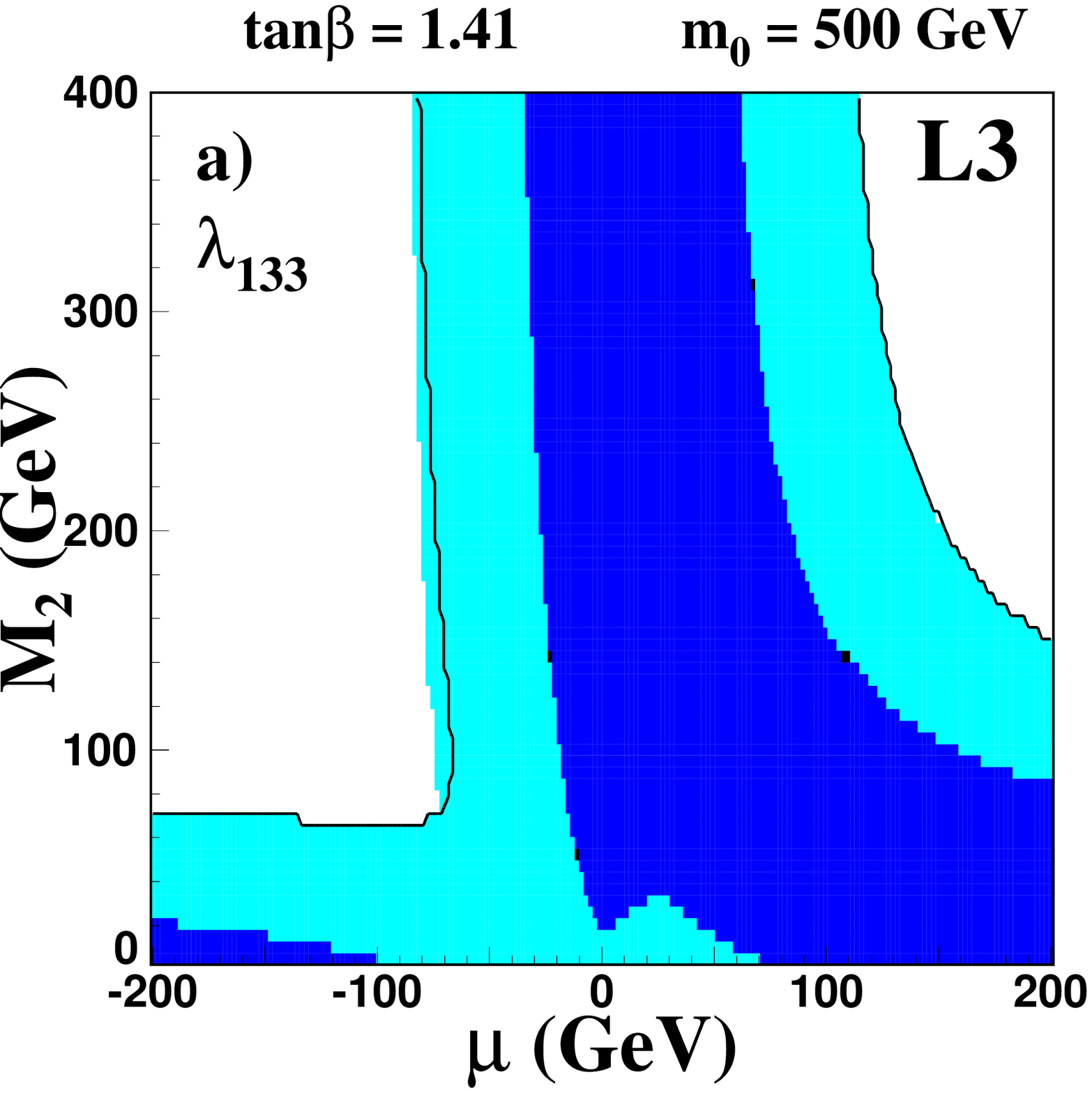}
  \end{center}
 \begin{center}
    \includegraphics[width=8 cm, height=8cm]{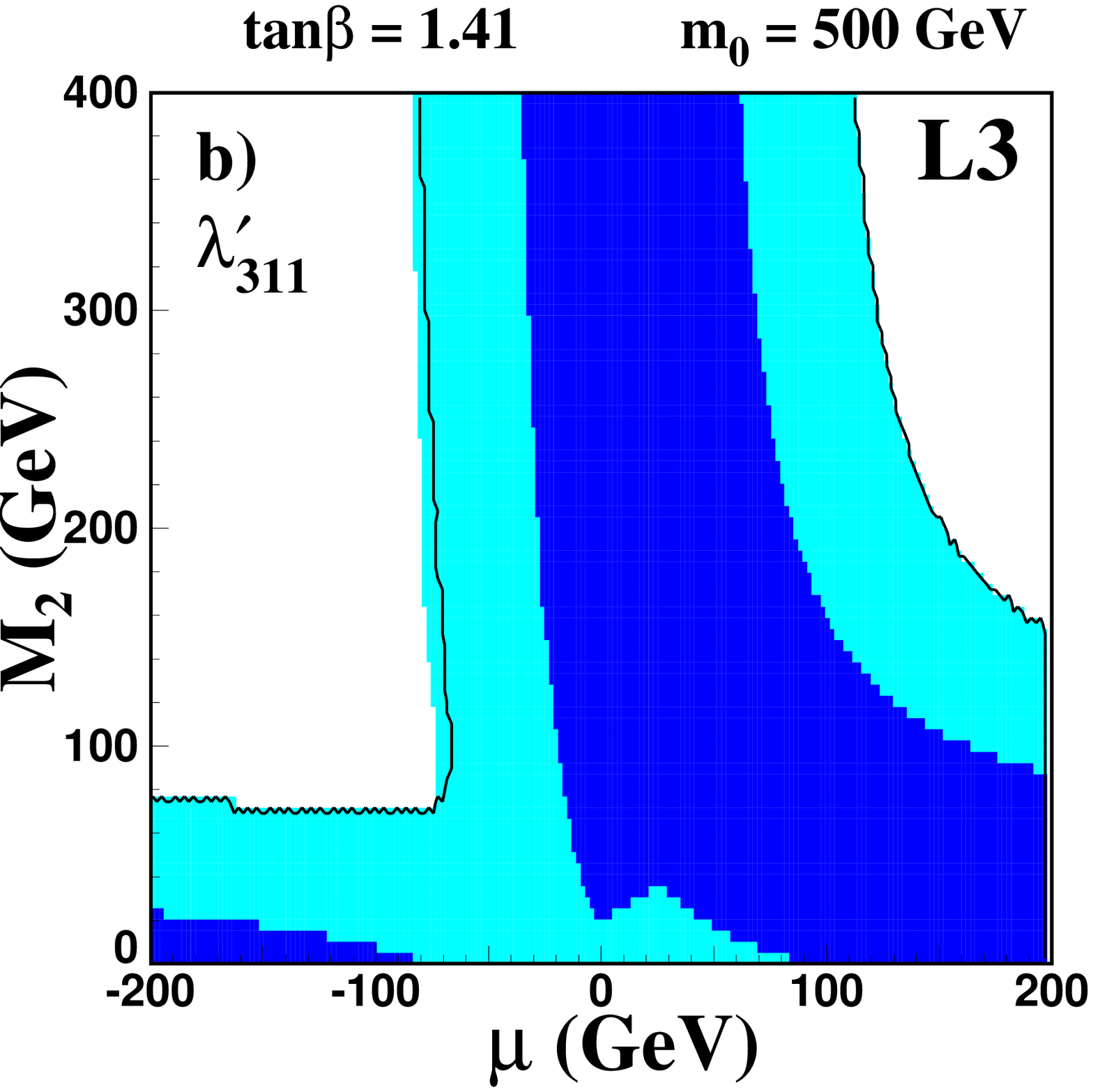}
    \includegraphics[width=8 cm, height=8cm]{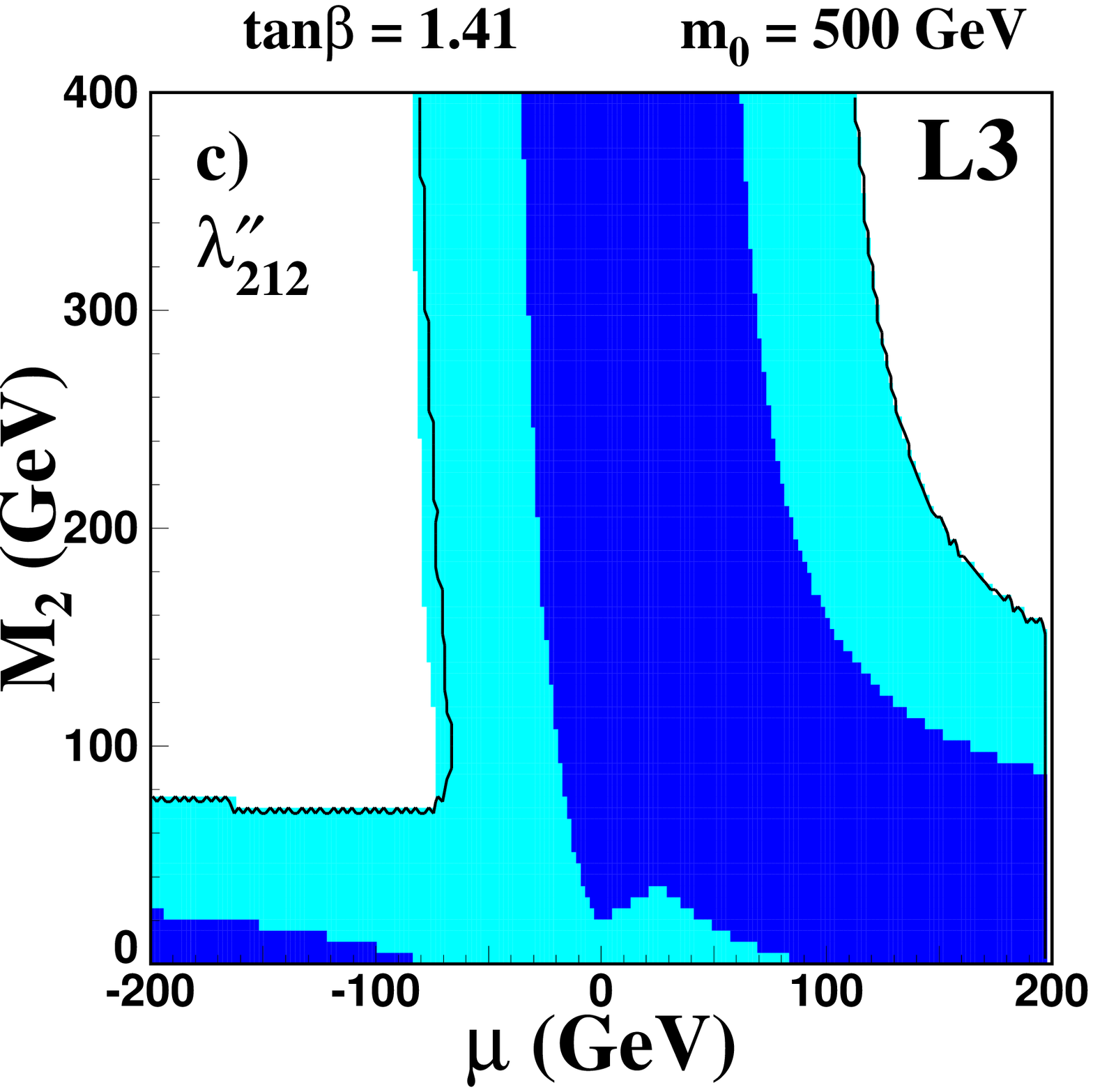}
  \end{center}
  \caption{Exclusion regions at 95\% C.L. for 
a) $\lamtre$, b) $\lambda'_{311}$  and c) $\lamuno$, for
$\tan\beta = \sqrt 2$ and ${m_0} = 500 $ \GeV.
The darker region is excluded by the Z lineshape measurements and the
lighter region by the present analyses. The black solid lines
indicate the chargino kinematic limit. The regions beyond
the kinematic limit are excluded by neutralino analyses. }
  \label{fig:excl_tot1}
\end{figure}

\begin{figure*}[htbp]  
\begin{center}
    \includegraphics[width=8 cm, height=8cm]{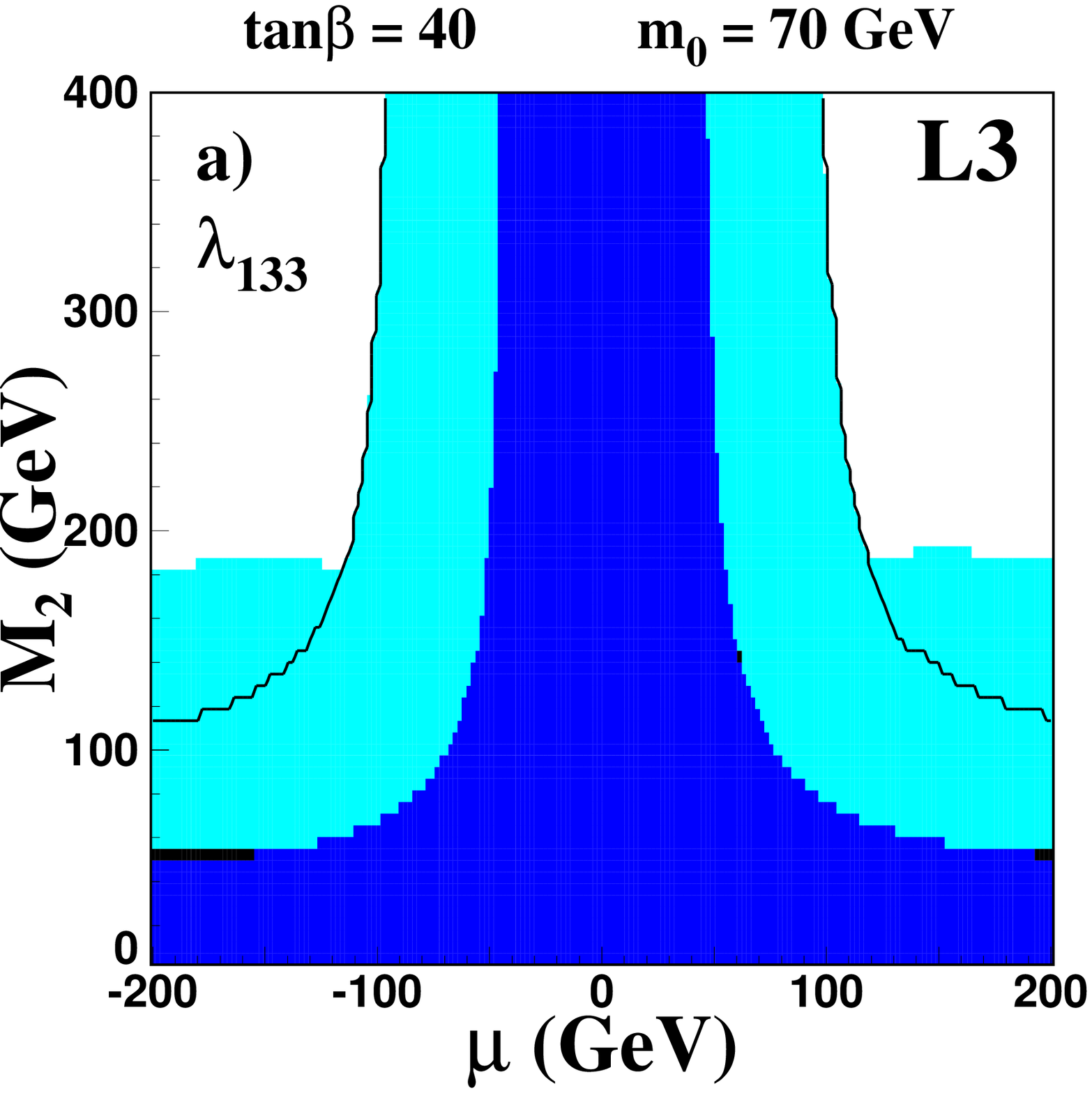}
  \end{center}
 \begin{center}
    \includegraphics[width=8 cm, height=8cm]{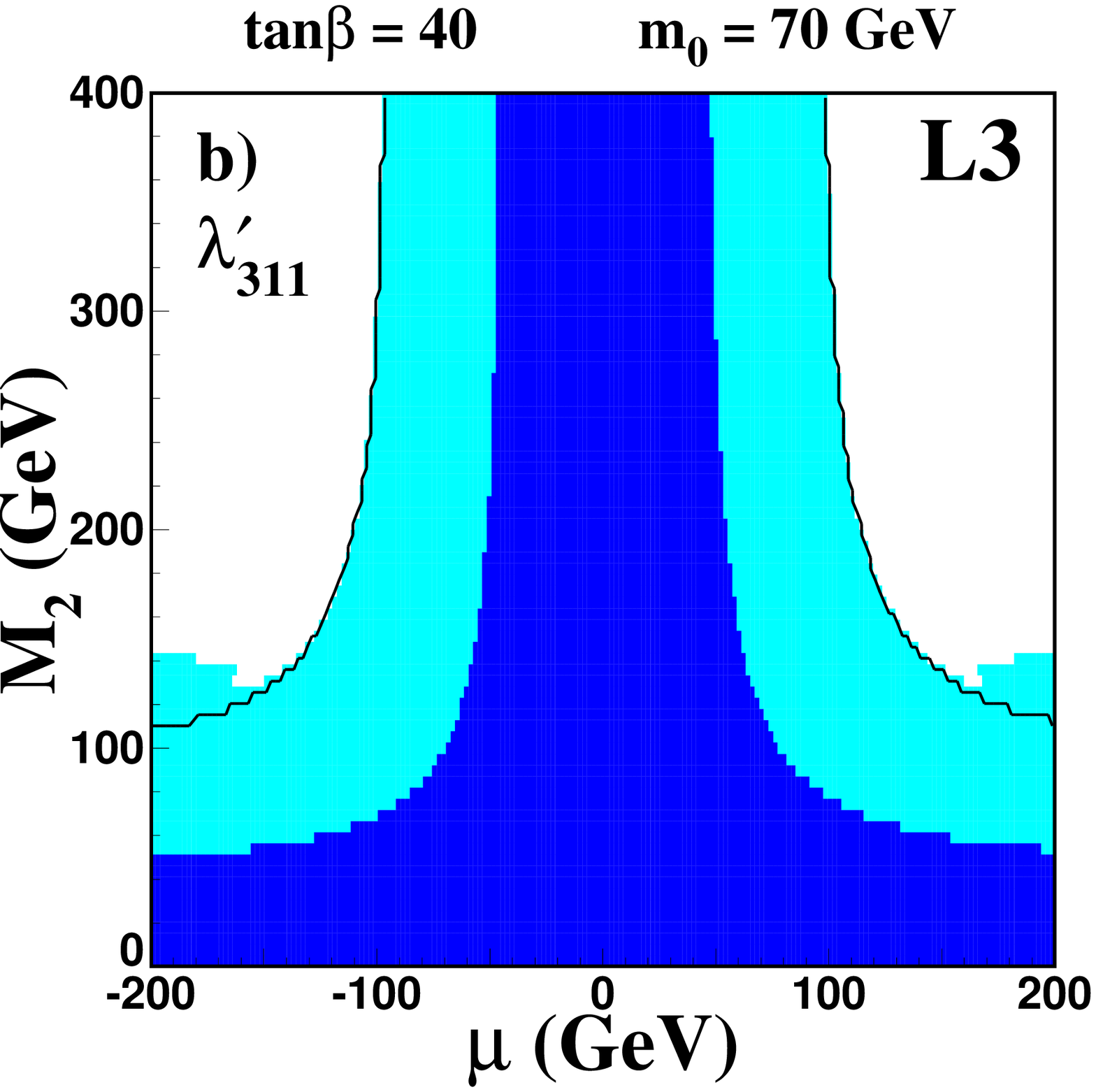}
    \includegraphics[width=8 cm, height=8cm]{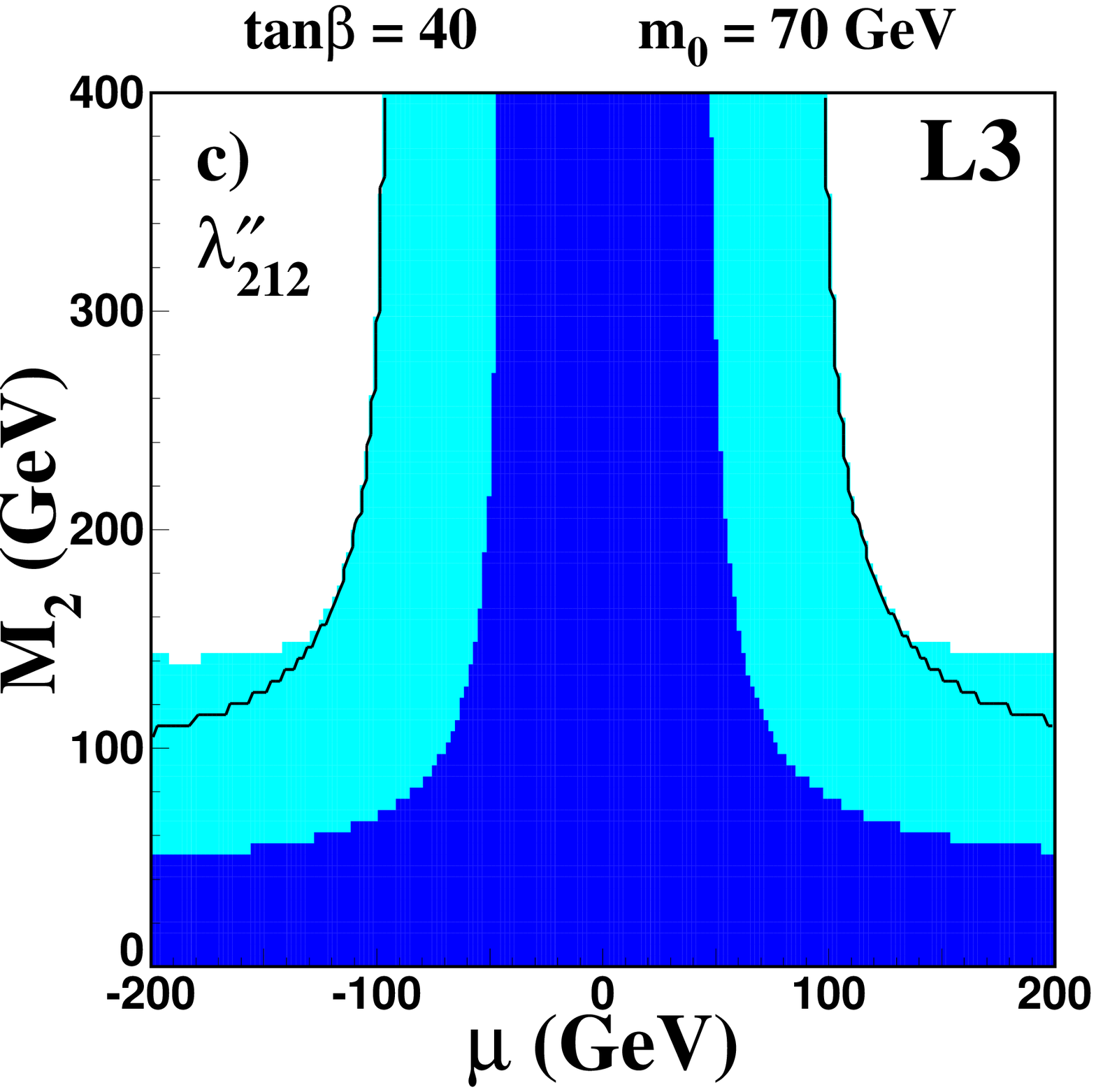}
  \end{center}
  \caption{
Exclusion regions at 95\% C.L. for 
a) $\lamtre$, b) $\lambda'_{311}$ and c) $\lamuno$, for
$\tan\beta = 40$ and ${m_0} = 70 $ \GeV.
The darker region is excluded by the Z lineshape measurements and the
lighter region by the present analyses. The black solid lines
indicate the chargino kinematic limit. The regions beyond
the kinematic limit are excluded by
neutralino analyses. }
  \label{fig:excl_tot2}
\end{figure*}

Figures~\ref{fig:excl_tot1} and~\ref{fig:excl_tot2} show the 
excluded regions at 95\% C.L. in the (${M_2, \mu}$) plane
for $\tan\beta=$ $\sqrt 2$ and ${m_0} = $ 500 \GeV,
and for  $\tan\beta=$ 40 and ${m_0} = $ 70 \GeV.
Some regions beyond the chargino
kinematic limit are excluded at large $m_0$  
and low $\tan\beta$ values by the 
$\chim\chin_{(n\geq 2)}$ analyses (Figure~\ref{fig:excl_tot1}) 
and at low $m_0$ by the $\chio\chio$ analyses (Figure~\ref{fig:excl_tot2}).

\subsection{\boldmath{Lightest scalar lepton $\slepr$ as LSP}}
\label{par:MSSM_slepton_LSP}

For 0 $\leq{m_0}<$ 50 \GeV{} and 1 $\leq \tan\beta<$2 the lightest
scalar lepton, the supersymmetric partner of the right-handed electron,
can be the LSP. In this region,
in the presence of dominant $\mathrm \lambda_{ijk}$ coupling, the
decay chain $\mathrm \chio \ra \ell \slepr \ra \ell \ell \nu$
leads to the same final states as those arising from neutralino
direct R-parity violating decays, so that the analysis is efficient
also when the lightest scalar lepton is the LSP.

When the $\mathrm \lambda'_{ijk}$ or $\mathrm \lambda''_{ijk}$
couplings dominate, the decays 
$\chio \ra \ell \slepr \ra \ell \ell \ell (\nu)$qq or
$\chio \ra \ell \slepr \ra \ell \ell$qqq  
occur, respectively. 
Those five-body decays are not implemented in 
{\tt SUSYGEN}. However, since in this region $\slepr$
is lighter than $\chio$, we can take into account the scalar lepton
decays $\slepr \ra \ell \ell (\nu)$qq  or
$\slepr \ra \ell$qqq, as described 
at the end of Sections~\ref{par:lambdapri_anal}
and~\ref{par:lambdasec_anal}.

\subsection{Mass Limits}
\label{par:mass_limits}

\begin{figure*}[htbp]  
\begin{center}
    \includegraphics[width=\figwidth]{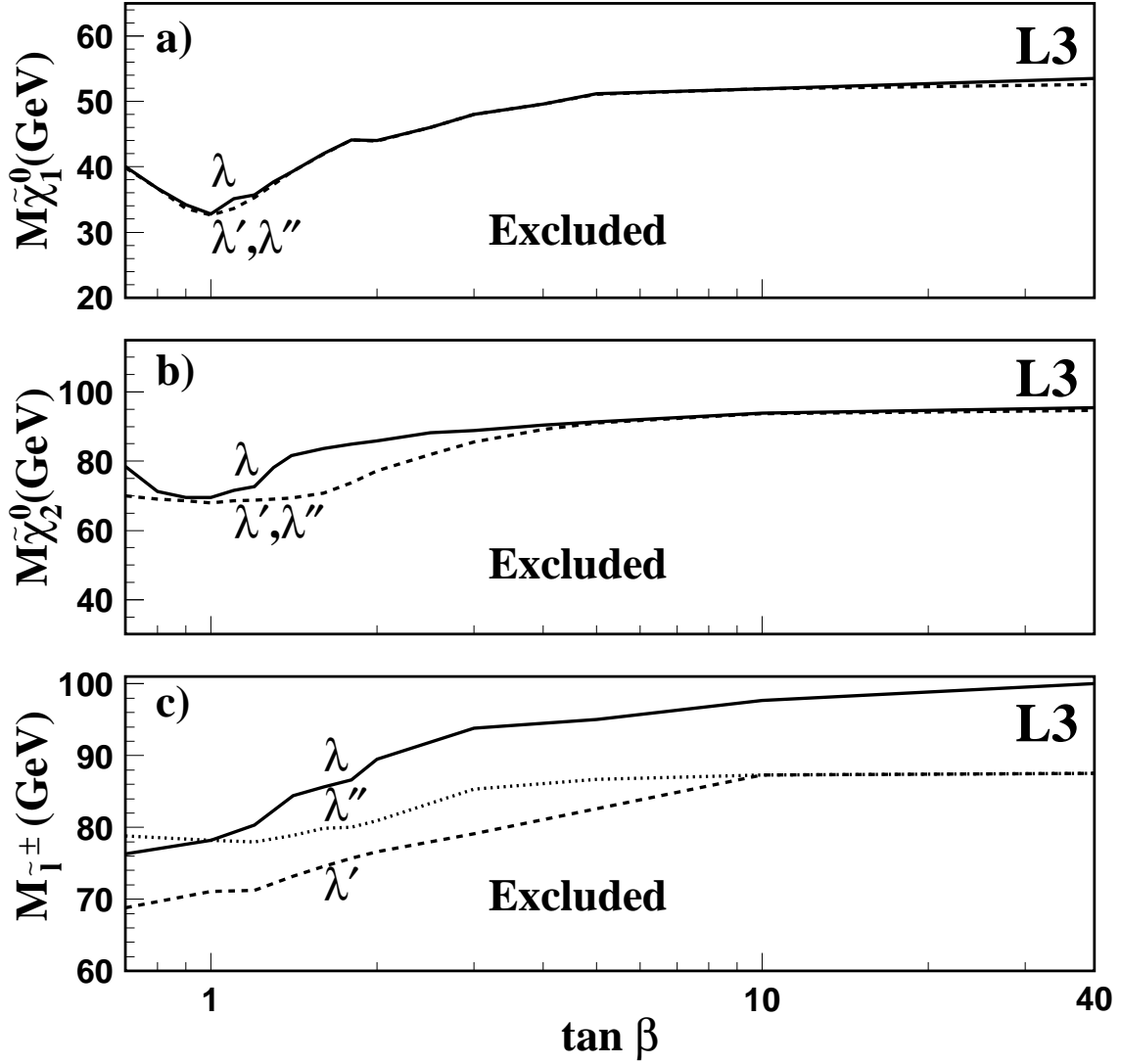}
  \end{center}
  \caption{
The solid, dashed and dotted lines, labelled with the corresponding
coupling, show the
95\% C.L. lower limits on the masses of a) $\protect\chio$, b) $\protect\chid$ and 
c) $\slepr$, as a function of \tb, for
$0 \leq M_2 \leq 1000$ \gev{} and  $- 500$ \GeV{} $\leq \mu \leq$ 500 \GeV.
$m_0 = $ 500 \gev{} in a) and b) and $m_0 = 0$ in c). }
  \label{fig:mlimit1}
\end{figure*}

\begin{figure*}[htbp]
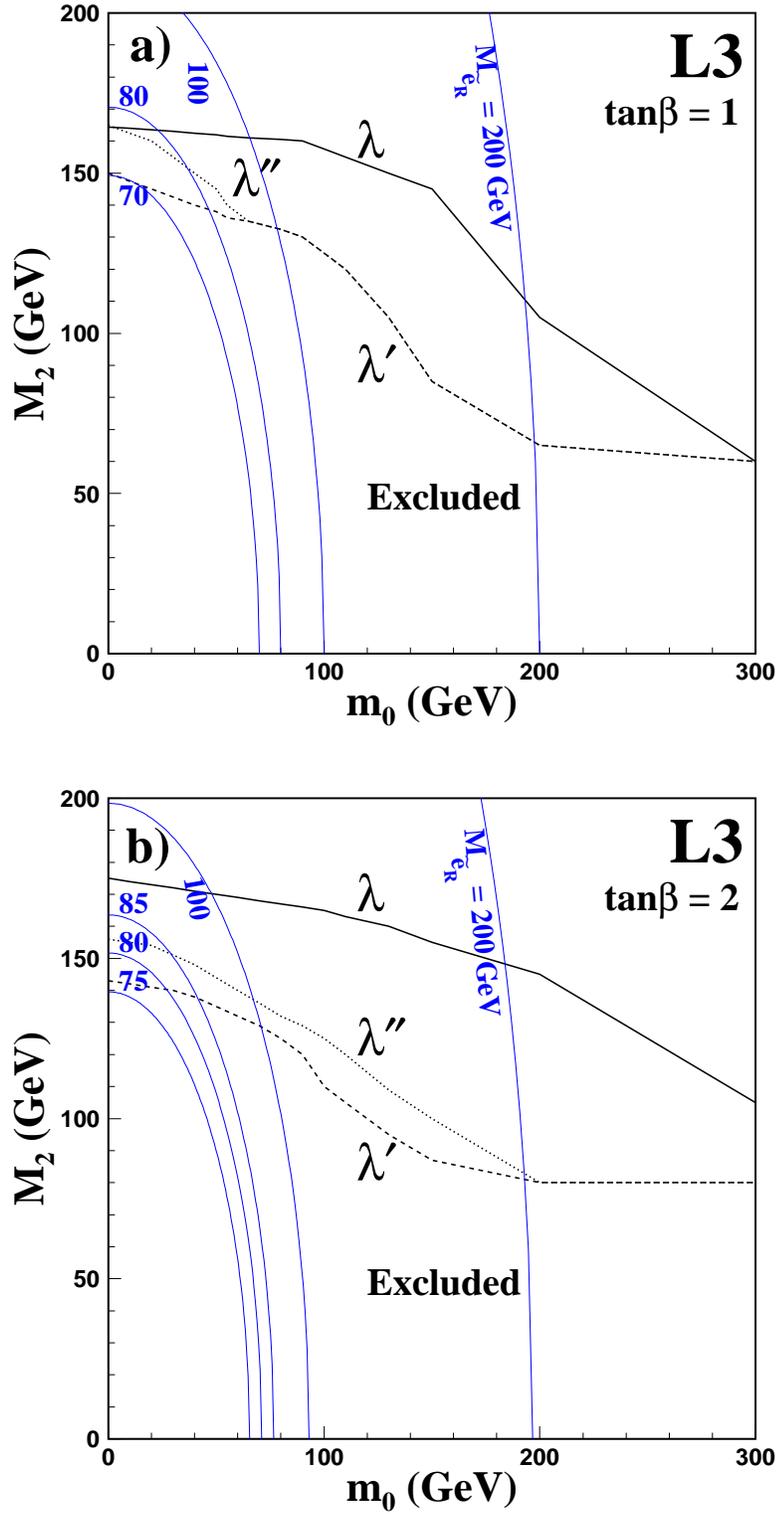
  
\begin{center}
    \includegraphics[width=9.5 cm, height=9 cm]{fig10a.epsi}
\end{center}
\vspace{0.5 cm}
\begin{center}
    \includegraphics[width=9.5 cm, height=9 cm]{fig10b.epsi}
  \end{center}
  \caption{
95\% C.L. exclusion contours in the ($M_2 \,$, $m_0$) plane, 
for $\mu < 0$ and  \tb $\,= 1$ (a) or \tb $\,= 2$ (b).
The lines labelled with the corresponding value in \gev{} represent the
contours of constant scalar electron mass $\mser$. 
The solid, dotted and dashed curves show the 95\% C.L. lower limits 
on $M_2$ as
a function of $m_0$. The contributions from scalar lepton, neutralino and
chargino searches are included. }
  \label{fig:m2m0}
\end{figure*}

Figure~\ref{fig:mlimit1} shows the 95\% C.L. lower limits on neutralino and
scalar lepton masses as a function of \tb. The $\chio$ and $\chid$ 
mass limits are shown for $m_0 = $ 500 \gev{} and the $\slepr$ ones for
$m_0 =$ 0.
At low \mo ($\mo \leq $ 70 \GeV), neutralino and scalar
lepton pair-production is enhanced, 
allowing to obtain better results also at low values of \tb.
For high \mo values, these contributions are suppressed and the mass limit 
is given by the chargino exclusion.
The absolute minimum on the scalar lepton mass is found at \mo = 0.
For low $m_0$ values the different 
contributions depend on \tb, since the $\slepr$
can be the LSP for low \tb\ values (\tb\ $<$ 2) and therefore only 
the scalar lepton analysis contributes to the limit in this region. 
For higher values of \tb, $\chio$ is the LSP, and the lower limit on the
scalar lepton mass is mainly given by the $\chio\chio$
exclusion contours. 
As an example,
Figure~\ref{fig:m2m0}a and~\ref{fig:m2m0}b  
show the 95\% C.L. lower limits on the mass
of the supersymmetric partner of the right-handed electron for \tb 
$\, = 1$ and 2, respectively.

\begin{figure}[htbp]
  \begin{center}
    \includegraphics[width=\figwidth]{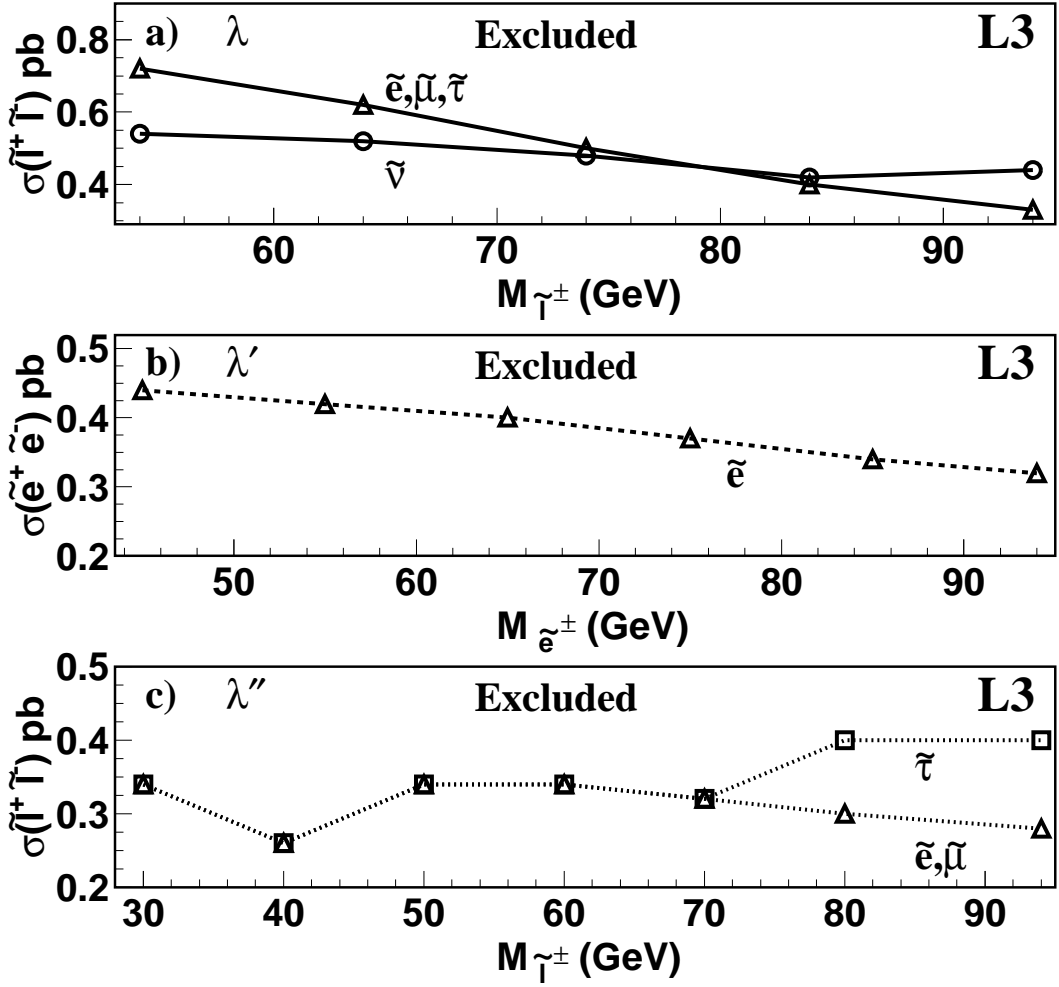}
  \end{center}
  \caption{
     95\% C.L. upper limits on the scalar lepton pair-production cross
    section, in the direct decay mode, as a function of the scalar lepton mass 
    for: a) $\lambda_{ijk}$, b) $\lambda'_{ijk}$ and c) $\lambda''_{ijk}$.
    The solid lines show the limits
    obtained by the $\lambda_{12k} $ analysis for scalar
    electrons, muons and taus and by the $\lambda_{121} $ 
    analysis for scalar neutrinos. The dashed line shows the limit obtained
    by the $\lambda'_{311}$ analysis for scalar
    electrons. The dotted lines show those obtained by 
    the $\lamuno$ analysis for scalar electrons, muons
    and taus.}
  \label{fig:cross_sec2}
\end{figure}

\begin{figure*}[htbp]  
\begin{center}
    \includegraphics[width=\figwidth]{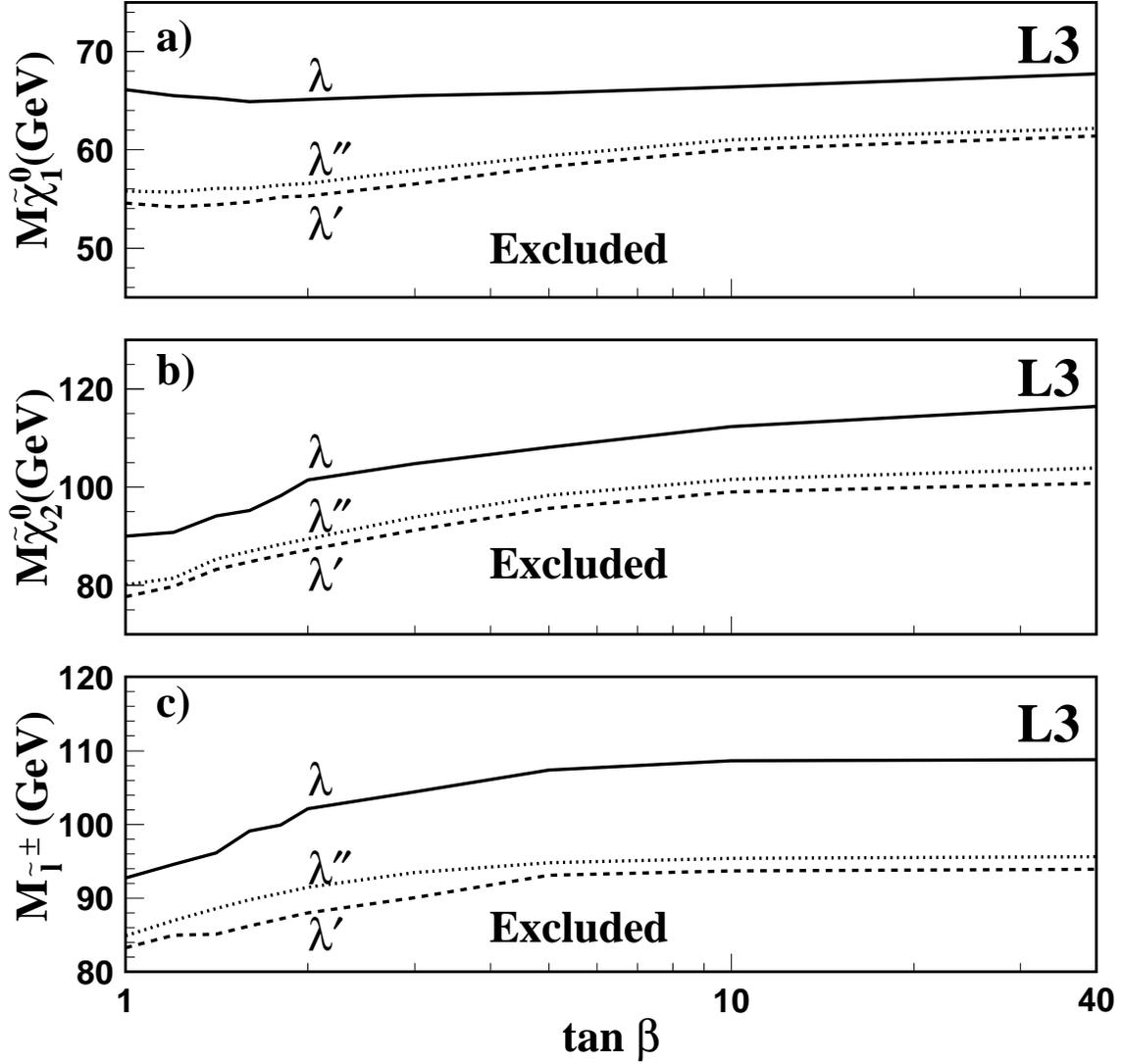}
  \end{center}
  \caption{
The solid, dashed and dotted lines, labelled with the corresponding
coupling, show the
95\% C.L. lower limits on the masses of a) $\protect\chio$, b) $\protect\chid$ and 
c) $\slepr$, as a function of \tb, for
$0 \leq M_2 \leq 1000$ \gev{},  $- 500$ \GeV{} $\leq \mu \leq$ 500 \GeV and
$m_0 = $ 50 \gev{}.}
  \label{fig:mlimit2}
\end{figure*}

The minima on the 95\% C.L. lower mass limits
shown in Figure~\ref{fig:mlimit1} correspond to the 
absolute minima from
the complete scan on $M_2$, $\mu$, ${m_0}$ and $\tan\beta$.
The absolute limit on $M_{\slepr}$ is found at $\tan\beta = 0.7$ in the
case of $\lambda_{ijk}$ and $\lambda'_{ijk}$ and at $\tan\beta = 1.4$ for
$\lambda''_{ijk}$. The difference is due to the lower 
cross-section upper limit
of $\lambda''_{ijk}$ for scalar lepton direct decays 
(Figure~\ref{fig:cross_sec2}), since for $\lambda_{ijk}$ and 
$\lambda'_{ijk}$ the limit on $M_{\slepr}$ is found when the $\slepr$
is the LSP. Figure~\ref{fig:mlimit2} shows similar mass limits
for $m_0 = $ 50 \gev.
The chargino mass limit is almost independent of $\tan\beta$,
being close to the kinematic limit for any value of $\tan\beta$ and $m_0$.

We derive lower limits at 95\% C.L. on the 
neutralino, chargino and scalar lepton masses, as detailed
in Table~\ref{tab:limits_189}. 
All analysis contributions (neutralino, chargino and scalar lepton
searches) are taken into account simultaneously
under the assumption of gaugino and scalar masses unification at the
GUT scale.

Identical scalar lepton mass limits are obtained even
without the  assumption of common scalar masses at the GUT scale.
For $\lambda_{ijk}$ and $\lambda'_{ijk}$ the present
bounds on the scalar lepton masses are found in the case in which
the $\slepr$  is the LSP. For $\lambda''_{ijk}$ this happens
when  the $\slepr$ and $\chio$ are nearly degenerate in mass. In both
cases
the neutralino analyses give the main contribution to the exclusion
in the regions of the parameter space around the limit.
The remaining sensitivity is due to searches for direct slepton decays
via $\lambda_{ijk}$. As these searches are equally sensitive to scalar
electron, muon or tau signals as shown in Figure~\ref{fig:cross_sec2},
the limits are unchanged.

\begin{table*} [htbp]
  \begin{center}
  \begin{tabular}{|c|c|c|c|} \hline 
 
 Particle Mass & $\lambda_{ijk}$ & $\lambda'_{ijk}$ & $\lambda''_{ijk}$  \\ \hline
  $\mchi$  \rpvpham    &  32.6 \gev     &   32.5 \gev     &  32.5 \gev      \\ \hline
  $\mchid$ \rpvpham    &  69.5 \gev     &   68.0 \gev     &  68.0 \gev      \\   \hline 
  $\mchit$ \rpvpham    &  99.3 \gev     &   99.0 \gev     &  99.0 \gev      \\ \hline
  $\mcha$  \rpvpham    &  94.3 \gev     &   93.8 \gev     &  93.8 \gev      \\ \hline
  $M_{\slepr}$ \rpvpham&  75.9 \gev     &   68.8 \gev     &  77.5 \gev      \\ \hline
  $M_{\snu}$  \rpvpham &  141.2 \gev    &  $-$            & $-$      \\ \hline

\end{tabular}
 \caption{Lower limits at 95\% C.L. on the masses of the supersymmetric
  particles considered in this analysis. The limits on $M_{\slepr}$ hold
  for ${\serr}$, $\smur$ and $\staur$.}
 \label{tab:limits_189}
\end{center}
\end{table*}

The search for R-parity violating decays of supersymmetric particles 
reaches at least the same sensitivity as in the R-parity 
conserving case~\cite{rpc_susy}.
Therefore, the supersymmetry limits obtained at LEP are independent of 
R-parity conservation assumptions.

\section*{Acknowledgements}

We wish to express our gratitude to the CERN accelerator divisions
for the excellent performance of the LEP machine. We acknowledge
the contributions of the engineers and technicians who have participated
in the construction and maintenance of this experiment.

%
\clearpage

\bibliographystyle{l3stylem}

\clearpage
%
\newpage
\typeout{   }     
\typeout{Using author list for paper 223 -- ? }
\typeout{$Modified: Tue Sep  5 19:04:46 2000 by clare $}
\typeout{!!!!  This should only be used with document option a4p!!!!}
\typeout{   }
%
%
%
%
%
%

\newcount\tutecount  \tutecount=0
\def\tutenum#1{\global\advance\tutecount by 1 \xdef#1{\the\tutecount}}
\def\tute#1{$^{#1}$}
\tutenum\aachen            
\tutenum\nikhef            
\tutenum\mich              
\tutenum\lapp              
\tutenum\basel             
\tutenum\lsu               
\tutenum\beijing           
\tutenum\berlin            
\tutenum\bologna           
\tutenum\tata              
\tutenum\ne                
\tutenum\bucharest         
\tutenum\budapest          
\tutenum\mit               
\tutenum\debrecen          
\tutenum\florence          
\tutenum\cern              
\tutenum\wl                
\tutenum\geneva            
\tutenum\hefei             
\tutenum\seft              
\tutenum\lausanne          
\tutenum\lecce             
\tutenum\lyon              
\tutenum\madrid            
\tutenum\milan             
\tutenum\moscow            
\tutenum\naples            
\tutenum\cyprus            
\tutenum\nymegen           
\tutenum\caltech           
\tutenum\perugia           
\tutenum\cmu               
\tutenum\prince            
\tutenum\rome              
\tutenum\peters            
\tutenum\potenza           
\tutenum\salerno           
\tutenum\ucsd              
\tutenum\santiago          
\tutenum\sofia             
\tutenum\korea             
\tutenum\alabama           
\tutenum\utrecht           
\tutenum\purdue            
\tutenum\psinst            
\tutenum\zeuthen           
\tutenum\eth               
\tutenum\hamburg           
\tutenum\taiwan            
\tutenum\tsinghua          

{
\parskip=0pt
\noindent
{\bf The L3 Collaboration:}
\ifx\selectfont\undefined
 \baselineskip=10.8pt
 \baselineskip\baselinestretch\baselineskip
 \normalbaselineskip\baselineskip
 \ixpt
\else
 \fontsize{9}{10.8pt}\selectfont
\fi
\medskip
\tolerance=10000
\hbadness=5000
\raggedright
\hsize=162truemm\hoffset=0mm
\def\r{\rlap,}
\noindent

M.Acciarri\r\tute\milan\
P.Achard\r\tute\geneva\ 
O.Adriani\r\tute{\florence}\ 
M.Aguilar-Benitez\r\tute\madrid\ 
J.Alcaraz\r\tute\madrid\ 
G.Alemanni\r\tute\lausanne\
J.Allaby\r\tute\cern\
A.Aloisio\r\tute\naples\ 
M.G.Alviggi\r\tute\naples\
G.Ambrosi\r\tute\geneva\
% ...
\newpage
\begin{list}{A}{\itemsep=0pt plus 0pt minus 0pt\parsep=0pt plus 0pt minus 0pt
                \topsep=0pt plus 0pt minus 0pt}
\item[\aachen]
 I. Physikalisches Institut, RWTH, D-52056 Aachen, FRG$^{\S}$\\
 III. Physikalisches Institut, RWTH, D-52056 Aachen, FRG$^{\S}$
\item[\nikhef] National Institute for High Energy Physics, NIKHEF, 
     and University of Amsterdam, NL-1009 DB Amsterdam, The Netherlands
\item[\mich] University of Michigan, Ann Arbor, MI 48109, USA
\item[\lapp] Laboratoire d'Annecy-le-Vieux de Physique des Particules, 
     LAPP,IN2P3-CNRS, BP 110, F-74941 Annecy-le-Vieux CEDEX, France
\item[\basel] Institute of Physics, University of Basel, CH-4056 Basel,
     Switzerland
\item[\lsu] Louisiana State University, Baton Rouge, LA 70803, USA
\item[\beijing] Institute of High Energy Physics, IHEP, 
  100039 Beijing, China$^{\triangle}$ 
\item[\berlin] Humboldt University, D-10099 Berlin, FRG$^{\S}$
\item[\bologna] University of Bologna and INFN-Sezione di Bologna, 
     I-40126 Bologna, Italy
\item[\tata] Tata Institute of Fundamental Research, Bombay 400 005, India
\item[\ne] Northeastern University, Boston, MA 02115, USA
\item[\bucharest] Institute of Atomic Physics and University of Bucharest,
     R-76900 Bucharest, Romania
\item[\budapest] Central Research Institute for Physics of the 
     Hungarian Academy of Sciences, H-1525 Budapest 114, Hungary$^{\ddag}$
\item[\mit] Massachusetts Institute of Technology, Cambridge, MA 02139, USA
\item[\debrecen] KLTE-ATOMKI, H-4010 Debrecen, Hungary$^\P$
\item[\florence] INFN Sezione di Firenze and University of Florence, 
     I-50125 Florence, Italy
\item[\cern] European Laboratory for Particle Physics, CERN, 
     CH-1211 Geneva 23, Switzerland
\item[\wl] World Laboratory, FBLJA  Project, CH-1211 Geneva 23, Switzerland
\item[\geneva] University of Geneva, CH-1211 Geneva 4, Switzerland
\item[\hefei] Chinese University of Science and Technology, USTC,
      Hefei, Anhui 230 029, China$^{\triangle}$
\item[\lausanne] University of Lausanne, CH-1015 Lausanne, Switzerland
\item[\lecce] INFN-Sezione di Lecce and Universit\`a Degli Studi di Lecce,
     I-73100 Lecce, Italy
\item[\lyon] Institut de Physique Nucl\'eaire de Lyon, 
     IN2P3-CNRS,Universit\'e Claude Bernard, 
     F-69622 Villeurbanne, France
\item[\madrid] Centro de Investigaciones Energ{\'e}ticas, 
     Medioambientales y Tecnolog{\'\i}cas, CIEMAT, E-28040 Madrid,
     Spain${\flat}$ 
\item[\milan] INFN-Sezione di Milano, I-20133 Milan, Italy
\item[\moscow] Institute of Theoretical and Experimental Physics, ITEP, 
     Moscow, Russia
\item[\naples] INFN-Sezione di Napoli and University of Naples, 
     I-80125 Naples, Italy
\item[\cyprus] Department of Natural Sciences, University of Cyprus,
     Nicosia, Cyprus
\item[\nymegen] University of Nijmegen and NIKHEF, 
     NL-6525 ED Nijmegen, The Netherlands
\item[\caltech] California Institute of Technology, Pasadena, CA 91125, USA
\item[\perugia] INFN-Sezione di Perugia and Universit\`a Degli 
     Studi di Perugia, I-06100 Perugia, Italy   
\item[\cmu] Carnegie Mellon University, Pittsburgh, PA 15213, USA
\item[\prince] Princeton University, Princeton, NJ 08544, USA
\item[\rome] INFN-Sezione di Roma and University of Rome, ``La Sapienza",
     I-00185 Rome, Italy
\item[\peters] Nuclear Physics Institute, St. Petersburg, Russia
\item[\potenza] INFN-Sezione di Napoli and University of Potenza, 
     I-85100 Potenza, Italy
\item[\salerno] University and INFN, Salerno, I-84100 Salerno, Italy
\item[\ucsd] University of California, San Diego, CA 92093, USA
\item[\santiago] Dept. de Fisica de Particulas Elementales, Univ. de Santiago,
     E-15706 Santiago de Compostela, Spain
\item[\sofia] Bulgarian Academy of Sciences, Central Lab.~of 
     Mechatronics and Instrumentation, BU-1113 Sofia, Bulgaria
\item[\korea]  Laboratory of High Energy Physics, 
     Kyungpook National University, 702-701 Taegu, Republic of Korea
\item[\alabama] University of Alabama, Tuscaloosa, AL 35486, USA
\item[\utrecht] Utrecht University and NIKHEF, NL-3584 CB Utrecht, 
     The Netherlands
\item[\purdue] Purdue University, West Lafayette, IN 47907, USA
\item[\psinst] Paul Scherrer Institut, PSI, CH-5232 Villigen, Switzerland
\item[\zeuthen] DESY, D-15738 Zeuthen, 
     FRG
\item[\eth] Eidgen\"ossische Technische Hochschule, ETH Z\"urich,
     CH-8093 Z\"urich, Switzerland
\item[\hamburg] University of Hamburg, D-22761 Hamburg, FRG
\item[\taiwan] National Central University, Chung-Li, Taiwan, China
\item[\tsinghua] Department of Physics, National Tsing Hua University,
      Taiwan, China
\item[\S]  Supported by the German Bundesministerium 
        f\"ur Bildung, Wissenschaft, Forschung und Technologie
\item[\ddag] Supported by the Hungarian OTKA fund under contract
numbers T019181, F023259 and T024011.
\item[\P] Also supported by the Hungarian OTKA fund under contract
  numbers T22238 and T026178.
\item[$\flat$] Supported also by the Comisi\'on Interministerial de Ciencia y 
        Tecnolog{\'\i}a.
\item[$\sharp$] Also supported by CONICET and Universidad Nacional de La Plata,
        CC 67, 1900 La Plata, Argentina.
\item[$\diamondsuit$] Also supported by Panjab University, Chandigarh-160014, 
        India.
\item[$\triangle$] Supported by the National Natural Science
  Foundation of China.
\end{list}
}
\vfill


\newpage

\clearpage

\newpage

\appendix
\section{Selection details}
\label{par:appendix_sel}

Dedicated selections are performed to maximize {\it a priori} the
analysis sensitivity $\mathrm 1/\sigma_{95}$, using
Monte Carlo signal and background events, where
$\mathrm \sigma_{95}$ is given by~\cite{grivaz}:
\begin{equation}
        \mathrm \sigma_{95} = \frac {1}{\varepsilon}
        \sum_{n = 0}^{\infty} c_n(b) P_b \,(n).
\end{equation}
Here $\mathrm \varepsilon$ is the signal selection efficiency, 
$\mathrm P_b \,(n) $ is the Poisson probability to observe
$n$ events with an expected background of $b$ events
and $\mathrm c_n(b)$ are the 95\% C.L. Bayesian
upper limits on the signal expectation values.

In Tables~\ref{tab:lambda_sel1} to~\ref{tab:lambdasec_sel3} 
we summarize all selections. 
In addition to the variables described in Sections~\ref{par:lambda_anal},
\ref{par:lambdapri_anal} and \ref{par:lambdasec_anal}, 
we apply cuts on the following variables:
\begin{itemize}

  \item $\lambda_{ijk}$: 
    \begin{itemize}
    \item event thrust,
    \item sum of the energies measured in the low angle calorimeters 
      (Luminosity monitor and Active Lead Ring)
      covering the polar angle region between $\mathrm 1.5^\circ$ and 
      $\mathrm 9.0^\circ$ ($E_{lum} + E_{alr}$),
    \item energy in the electromagnetic ($E_{bgo}$) and hadronic
      calorimeters ($E_{hcal}$),
    \item energy ($E^{\ell}_n$) of the
      $n$ identified leptons.
    \end{itemize}

   \item $\lambda'_{ijk}$: 
     \begin{itemize}
     \item energy in a cone of $\mathrm 12^\circ$ half-opening angle around the
       beam axis ($E_{v12}$), 
     \item sphericity,  
     \item probability 
       ($Prob(\chi^2_{WW},5)$ or $Prob(\chi^2_{ZZ},5)$) that the
       reconstructed invariant mass, after a 5C fit, is consistent with 
       $\WpWm$ or ZZ pair-production,
     \item energy ($E_{20}^{\ell}$) in a cone of $\mathrm 20^\circ$ 
       half-opening 
       angle around the direction of the lepton candidate, calculated
       subtracting the lepton energy,
     \item jet invariant masses ($M_{jet1}$ and $M_{jet2}$).
     \end{itemize}

  \item $\lambda''_{ijk}$: 
    \begin{itemize}
   \item jet widths
   \item $E_{20}^{\ell}$. 
\end{itemize}
\end{itemize}

The cut values are chosen 
according to the procedure described above.
In particular, the selection criteria for the variables marked with~* 
are optimized simultaneously.

\begin{table*}[hpb] 
  \begin{center}
  \begin{tabular}{|l|c|c|c|c|} \hline 

    \multicolumn{5}{|c|} {$\lambda_{ijk}: \;
    4\ell + \Emiss \,$ } \cr \hline

   \multicolumn{1}{|c|} { }
   &\multicolumn{2}{c|} {Low mass values $( \leq 25 \gev)$ }
   &\multicolumn{2}{c|} {High mass values $( \geq 25 \gev)$ } \cr \hline

    &$\ell =$ e, $\mu$ only & $\ell = \tau $ &$\ell =$ e, $\mu$ only & $\ell = \tau $           \\ \hline
   $N_{e,\mu}$      & $\geq$ 2 &   $-$             & $\geq$ 2 & $-$                 \\ \hline
   $N_{e,\mu,\tau}$ & $-$ & $\geq$ 3 &  $-$            &  $\geq$ 3
    \\ \hline
   $N_{tracks}$     & 4 & 4 -- 6   &  3 -- 4         &  3 -- 8                      \\ \hline
   $N_{clusters}$   &  4 -- 8 & 4 -- 16  &  4 -- 8         &  6 -- 24               \\ \hline
   $E_{lum}+ E_{alr}$ (\gev)& $<$ 10  & $<$ 10  &  $<$ 10    & $<$ 10              \\ \hline 
   $E_{bgo}$ (\gev) & $<$ 150 &  $>$ 20   &  $<$ 100     & $-$                    \\ \hline 
   $E_{hcal}$ (\gev)& $<$ 10  & $<$ 15  &  $<$ 10    & $-$              \\ \hline 
   ${E_{vis}}/{\sqrt{s}}$ & 0.20 -- 0.90 & 0.20 -- 0.60& 0.20 -- 0.90 & 0.20 -- 0.60    \\ \hline
   $p^T_{miss}$ (\gev)      & $>$ 7  & $>$ 7 & $>$ 7 & $>$ 7         \\ \hline
 * ${\it Thrust}$        & $-$  & $-$  &  $<$ 0.947  & $<$ 0.918                       \\ \hline
 * sin$(\theta_{miss})$     & $>$ 0.277 & $>$ 0.486 & $>$ 0.430 & $>$ 0.436 \\ \hline
 * $\theta_{acol}$ (rad)    & $<$ 3.107 & $<$ 3.135 & $<$ 3.035 & $<$ 2.952 \\ \hline
 * $\theta_{acop}$ (rad)    & $<$ 3.107 & $-$ & $<$ 3.050 & $<$ 3.044    \\ \hline 
 * $\ytq         $  & 0.0002 -- 0.0018 & 0.0002 -- 0.0020 & $>$ 0.00060$$ & $>$ 0.00056$$     \\ \hline 

  \end{tabular}                     
  \caption{Cut values of the $4\ell + \Emiss \,$ selections, for final
   states with at least one $\tau$ or with $\ell =$ e, $\mu$ only. 
   Topologies with four leptons plus missing energy result from neutralino, 
   chargino and scalar neutrino decays, 
   as detailed in Table~\ref{tab:topologies}. 
   Different selection criteria are developed according to the
   mass values of the pair-produced supersymmetric particles. }
  \label{tab:lambda_sel1}
  \end{center}
\end{table*}

\begin{table*} 
  \begin{center}
  \begin{tabular}{|l|c|c|} \hline 

    \multicolumn{3}{|c|} {$\lambda_{ijk}: \;
    2\ell + \Emiss \,$ and $\, 6 \ell \,$} \cr \hline

   \multicolumn{1}{|c|} { }
   &\multicolumn{1}{c|} {$2 \ell + \Emiss \,$ }
   &\multicolumn{1}{c|} {$6 \ell$ } \cr \hline

    &$\ell =$ e, $\mu, \tau $    &$\ell =$ e, $\mu, \tau $ \\ \hline

   $N_{e,\mu,\tau}$          &  $ =$ 2      & $\geq$ 4     \\ \hline
   $N_{tracks}$              &  2 -- 4      &  5 -- 11     \\ \hline
   $N_{clusters}$            &  3 -- 5      &  6 -- 18     \\ \hline
   $N_{jets8}$               & $ =$ 2       & $-$          \\ \hline 
   $E_{lum}+ E_{alr}$ (\gev) & $<$ 10       &  $<$ 10      \\ \hline 
   $E_{bgo}$ (\gev)          & $-$          &  15 -- 125   \\ \hline 
   $E_1^{\ell}$ (\gev)       & 30 -- 60     & $-$          \\ \hline 
   $E_1^{\ell} + E_2^{\ell}$(\gev)& $>$ 40  & $-$          \\ \hline 
   ${E_{vis}}/{\sqrt{s}}$    & 0.25 -- 0.55 & 0.30 -- 1.20 \\ \hline
   $p^T_{miss}$ (\gev)       & $>$ 7        & $>$ 7        \\ \hline
 * ${\it Thrust}$            & $-$          & $<$ 0.971    \\ \hline
 * sin$(\theta_{miss})$      & $>$ 0.278    & $>$ 0.314    \\ \hline
 * $\theta_{acol}$ (rad)     & $<$ 2.903    & $-$          \\ \hline 
 * $\theta_{acop}$ (rad)     & $<$ 2.904    & $-$          \\ \hline 
 * $\ytq         $           & $-$          & $>$ 0.0009   \\ \hline 

  \end{tabular}                     
  \caption{Selection criteria of the $2\ell + \Emiss \,$ and $6 \ell$
   selections.
   Topologies with two leptons plus missing energy arise from 
   chargino and scalar lepton direct decays, as shown in 
   Table~\ref{tab:topologies}.
   Six lepton topologies result from chargino direct decays.
   Inclusive selection criteria 
   are developed for final states with $\ell =$ e, $\mu$ or $ \tau$.}
  \label{tab:lambda_sel2}
  \end{center}
\end{table*}

\begin{table*} 
  \begin{center}
\begin{sideways}
\begin{minipage}[b]{\textheight}
\begin{center}

  \begin{tabular}{|l|c|c|c|c|c|c|c|} \hline 
    \multicolumn{8}{|c|} 
  {$\lambda_{ijk}: \; (\geq 4) \,\ell \, +$ (jets) $+ \Emiss \,$, 
     with $\ell =$ e, $\mu$ only} \cr \hline

     & $\DM \leq $ 20 \gev & \multicolumn{3}{c|}{$\DM = $ 20 -- 50 \gev}  
                         & \multicolumn{3}{c|}{$\DM \geq $ 50 \gev} \\ \hline 
   $N_{e,\mu}$      & $\geq$ 2 & \multicolumn{3}{c|} {$\geq$ 2}
                               & \multicolumn{3}{c|}{$\geq$ 2} \\ \hline
   $E_{lum}+E_{alr}$ (\gev)& $<$ 10 & \multicolumn{3}{c|} {$<$ 10 }
                               & \multicolumn{3}{c|}{$<$ 10 } \\ \hline
   $E_{bgo}$ (\gev) & $<$ 110  & \multicolumn{3}{c|} {$<$ 110 }
                               & \multicolumn{3}{c|}{$<$ 110 } \\ \hline
   $p^T_{miss}$ (\gev)         & $>$ 7  & \multicolumn{3}{c|} {$>$ 7 }
                               & \multicolumn{3}{c|}{$>$ 7} \\ \hline 
  * ${\it Thrust}$          & $<$ 0.961 & \multicolumn{3}{c|} {$<$0.878} 
                               & \multicolumn{3}{c|}{$<$0.811}  \\ \hline
  * sin($\theta_{miss}$)  & $>$ 0.259 & \multicolumn{3}{c|}{$>$ 0.362}
                        & \multicolumn{3}{c|}{$>$ 0.389} \\ \hline
  * $\theta_{acol}$ (rad) & $<$ 3.044 & \multicolumn{3}{c|}{ $<$ 3.112} 
                               & \multicolumn{3}{c|}{ $<$ 2.998} \\ \hline
  * $\theta_{acop}$ (rad) & $<$ 3.139 & \multicolumn{3}{c|}{ $<$ 3.136}
                               & \multicolumn{3}{c|}{ $<$ 3.065} \\ \hline
  * $\ytq         $        & $>$ 0.0026 & \multicolumn{3}{c|}{ $>$ 0.0109 }
                               & \multicolumn{3}{c|}{ $>$ 0.0118 } \\ \hline
    &  & $\mathrm{hadronic}$ & $\mathrm{mixed}$ & $\mathrm{leptonic}$ 
       & $\mathrm{hadronic}$ & $\mathrm{mixed}$ & $\mathrm{leptonic}$ \\ \hline
    $N_{tracks}$ & 5 -- 11 & 12 -- 26 & 8 -- 17 & 5 -- 8 
                     & 18 -- 28 & 8 -- 22 & 5 -- 6 \\ \hline
    $N_{clusters}$ & 6 -- 26 & 35 -- 52 & 18 -- 40 & 6 -- 12 
                     & 38 -- 60 & 20 -- 42 & 6 -- 11 \\ \hline
    ${E_{vis}}/{\sqrt{s}}$ & 0.30--0.90  
          & 0.50--0.85  & 0.40--0.85  & 0.35--0.75 
          & 0.60--0.90  & 0.45--0.80  & 0.30--0.75 \\ \hline
  \end{tabular}                        
\end{center}
  \caption{Cut values of the $(\geq 4) \,\ell + $ (jets) $ + \Emiss \,$ selections, 
   for final states with $\ell =$ e, $\mu$ only. 
   Topologies with multileptons plus possible jets and missing energy
   result from neutralino, chargino and scalar lepton indirect decays,
   as presented in Table~\ref{tab:topologies}. 
   Different selection criteria are developed depending on the
   mass differences 
   $\DM = \protect\mcha - \protect\mchi$, $\protect\mchin - \protect\mchi$ or $\mslepr - \protect\mchi$.
   For large and medium $\DM$ three selections are applied according to 
   the virtual W-pair decays into hadronic, mixed or leptonic final states.}
  \label{tab:lambda_sel3a}
\end{minipage}
\end{sideways}
  \end{center}
\end{table*}

\newpage
\begin{table*} 
  \begin{center}
  \begin{tabular}{|l|c|c|c|} \hline
    \multicolumn{4}{|c|} 
  {$\lambda_{ijk}: \; (\geq 4) \, \ell +$ (jets) $+ \Emiss \,$,
     with $\ell = \tau$} \cr \hline
 
    & $\DM \leq $ 20 \gev & \multicolumn{2}{c|}{$\DM \geq $ 20 \gev}  
                         \\ \hline 
    $N_{e,\mu,\tau}$     & $\geq$ 3 & \multicolumn{2}{c|}{$\geq$ 3}     \\ \hline
    $E_{lum}+E_{alr}$ (\gev)& $<$ 10 & \multicolumn{2}{c|} {$<$ 10 }    \\ \hline
    $p^T_{miss}$ (\gev)     & $>$ 7  & \multicolumn{2}{c|} {$>$ 7 }     \\ \hline
  * ${\it Thrust}$        & $<$ 0.991 & \multicolumn{2}{c|}{$<$0.832}       \\ \hline
  * sin($\theta_{miss}$)  & $>$ 0.396 & \multicolumn{2}{c|}{$>$ 0.464}  \\ \hline
  * $\theta_{acol}$ (rad) & $<$ 3.058 & \multicolumn{2}{c|}{$<$ 3.092}  \\ \hline
  * $\theta_{acop}$ (rad) & $-$       & \multicolumn{2}{c|}{$<$ 3.140}  \\ \hline
  * $\ytq         $       &$>$ 0.0018 & \multicolumn{2}{c|}{$>$ 0.0141} \\ \hline
    &  & $\mathrm{hadronic}$ $\mathrm{and}$  $\mathrm{mixed}$ & 
                                             $\mathrm{leptonic}$  \\ \hline
    $N_{tracks}$ &   6 -- 12 & 8 -- 25    & 5 -- 12  \\ \hline
    $N_{clusters}$ & 6 -- 32 & 20 -- 52   & 8 -- 24 \\ \hline
    ${E_{vis}}/{\sqrt{s}}$ & 0.30--0.65 
                  & 0.30--0.70 & 0.30--0.70 \\ \hline
  \end{tabular}
  \caption{Selection criteria 
   of the $(\geq 4) \,\ell + $ (jets) $ + \Emiss \,$ selections, 
   for final states with at least one $\tau$. 
   Topologies with multileptons plus possible jets and missing energy
   arise from neutralino, chargino and scalar lepton indirect decays,
   as listed in Table~\ref{tab:topologies}. 
   Different selection criteria are developed depending on the
   mass differences $\DM = \protect\mcha - \protect\mchi$, $\protect\mchin - \protect\mchi$ or $\mslepr - \protect\mchi$.
   For $\DM$ greater than 20 \gev{}, three selections are applied according to 
   the virtual W-pair decays into hadronic and mixed or leptonic final states.}
  \label{tab:lambda_sel3b}
  \end{center}
\end{table*}


\begin{table*}
\begin{center}
\begin{tabular}{|l|c|c|c|}\hline

        \multicolumn{4}{|c|} {$\lambda'_{ijk}: \;$
 4 jets $\, + \,2 \tau$ }\cr\hline

     & Low masses       & Medium masses   & High masses \\
     & ($\leq$ 30 \gev) & (35--50 \gev)   & ($\geq$ 55 \gev) \\\hline
$E_{v12}/E_{vis}$ & \multicolumn{3}{c|}{$ <$ 0.20} \\ \hline
$E_{lum}+E_{alr}$ (\gev) & \multicolumn{3}{c|}{$ <$ 1 } \\ \hline
W mass window & \multicolumn{3}{c|}{reject if \; 70 \gev $< M_{qq}<$  95 \gev 
             \; and 70 \gev $< M_{\ell\nu} <$ 90 \gev}  \\\hline
$ N_{tracks}$   & 8--29 & 5--34 & 13--38 \\\hline
$N_{clusters}$  & 16--63 & 24--73 & 43--94 \\\hline
$N_{jets8}$      & -- & -- & $\geq 4$ \\\hline
$N_{e,\mu,\tau}$ & $\geq 1$ & $\geq 1$ & $\geq 2$ \\\hline
$E_{miss}/\sqrt{s}$ & $<0.33$ & $<0.35$ & $<0.24$ \\\hline
${\it Prob}(\chi^2_{WW},5)$ & -- & $<0.5$ & $<0.3$ \\\hline
${\it Prob}(\chi^2_{ZZ},5)$ & -- & -- & $<0.3$ \\\hline
${\it Sphericity}$ & 0.003--0.3 & $<0.42$ & 0.1--0.81 \\\hline
$W_{jet1}$ & 0.04--0.36 & $\geq 0.1$ & $\geq 0.1$ \\\hline
$W_{jet2}$ & 0.04--0.36 & $\geq 0.1$ &  -- \\\hline
$|M_{jet1}-M_{jet2}|$ (\gev) & $<$24 & -- & -- \\\hline 
$\theta_{acol}$ (rad) & -- & $<$3.09 & -- \\\hline
$E_{bgo}/\sqrt{s}$ & 0.1--0.7 & 0.2--0.6 & 0.2--0.6\\\hline
$E_{hcal}/\sqrt{s}$ & 0.05--0.4 & 0.05--0.3 & 0.05--0.35\\\hline

* ${\it Thrust}$ & 0.94--0.99 & 0.83--0.98 & 0.61--0.83 \\\hline
* ln($\ytq$)            & $>-6.18 $&$>-5.55$ & $>-4.73$ \\\hline
* sin($\theta_{miss}$)  & $>0.24$ & $>0.67$ & $>0.66$ \\\hline
* $E_{vis}/{\sqrt{s}}$  & 0.59--0.89 & 0.67--0.95 & 0.69--0.88 \\\hline
* $\theta_{acop}$ (rad) & $<$ 3.139 & $<$3.093 & $<$ 3.136 \\\hline
\end{tabular}
\caption{Values of the selection requirements for 
the ``4 jets + 2 taus'' selections.
These final states arise from neutralino decays via 
$\lambda'$ coupling, as shown in Table \ref{tab:topologies_lqd}.}
\label{tab:lambdaprime_sel1}
\end{center}
\end{table*}

\begin{table*}
\begin{center}
\begin{tabular}{|l|c|c|c|}\hline

        \multicolumn{4}{|c|} {$\lambda'_{ijk}: \;$
 4 jets $\, + \,\tau + \Emiss$ }\cr\hline
     & Low masses       & Medium masses   & High masses \\
     & ($\leq$ 30 \gev) & (35--50 \gev)   & ($\geq$ 55 \gev) \\\hline
$E_{v12}/E_{vis}$ & \multicolumn{3}{c|}{$ <$ 0.20} \\ \hline
$E_{lum}+E_{alr}$ (\gev) & \multicolumn{3}{c|}{$ <$ 1 } \\ \hline
$N_{e,\mu,\tau}$ & \multicolumn{3}{c|}{$\geq$ 1}\\\hline
W mass window & \multicolumn{3}{c|}{reject if \; 70 \gev $< M_{qq}<$  95 \gev 
             \; and 70 \gev $< M_{\ell\nu} <$ 90 \gev}  \\\hline
$ N_{tracks}$ & 5--29 & 6--38 & 19--42 \\\hline
$N_{clusters}$ & 15--65 & 20--77 & 43--97 \\\hline
$E_{miss}/\sqrt{s}$ & $<$0.37 & $<$0.42 & 0.08--0.34 \\\hline
${\it Prob}(\chi^2_{WW},5)$ & $<$0.8 & $<$0.16 & $<$0.17 \\\hline
${\it Prob}(\chi^2_{ZZ},5)$ & $<$0.4 & -- & $<$0.5 \\\hline
${\it Sphericity}$ & $<$0.2 & $<$0.46 & 0.08--0.8 \\\hline
$W_{jet1}$ & 0.03--0.4 & 0.04--0.65 & $\geq 0.2$ \\\hline
$W_{jet2}$ & 0.03--0.5 & 0.04--0.65 & $\geq 0.05$ \\\hline
$|M_{jet1}-M_{jet2}|$ (\gev) & $<$28 & -- & -- \\\hline 
$E_{bgo}/\sqrt{s}$ & 0.1--0.66 & 0.13--0.66 & 0.21--0.56\\\hline
$E_{hcal}/\sqrt{s}$ & 0.03--0.41 & 0.03--0.41 & $<$0.40\\\hline

* ${\it Thrust}$ & 0.95--0.99 & 0.84--0.99 & 0.60--0.85 \\\hline
* ln($\ytq$) & $> -7.54$ & $> -6.74$ & $> -4.89$ \\\hline
* sin($\theta_{miss}$) & $>$0.35 & $>$0.68 & $>$0.47 \\\hline
* $E_{vis}/{\sqrt{s}}$& 0.63--0.92 & 0.67--0.84 & 0.62--0.82 \\\hline
* $\theta_{acop}$ (rad) & $<$3.05 & $<$3.02 & -- \\\hline
\end{tabular}
\caption{Cut values of the ``4 jets $+ \,\tau + \Emiss$'' selections.
These final states arise from neutralino decays via 
$\lambda'$ coupling, as detailed in 
Table~\ref{tab:topologies_lqd}.}
\label{tab:lambdaprime_sel2}
\end{center}
\end{table*}

\begin{table*}
\begin{center}
\begin{tabular}{|l|c|c|c|}\hline

        \multicolumn{4}{|c|} {$\lambda'_{ijk}: \;$
 4 jets $ \,+ \,\Emiss$ }\cr\hline
     & Low masses       & Medium masses   & High masses \\
     & ($\leq$ 30 \gev) & (35--50 \gev)   & ($\geq$ 55 \gev) \\\hline
$E_{v12}/E_{vis}$ & \multicolumn{3}{c|}{$ <$ 0.20} \\ \hline
$E_{lum}+E_{alr}$ (\gev) & \multicolumn{3}{c|}{$ <$ 1 } \\ \hline
$N_{jets8}$      & -- & -- & $\geq 3$ \\\hline
$N_{e,\mu,\tau}$ & \multicolumn{3}{c|}{$\geq$ 1}\\\hline
W mass window & \multicolumn{3}{c|}{reject if \; 70 \gev $< M_{qq}<$  95 \gev 
             \; and 70 \gev $< M_{\ell\nu} <$ 90 \gev}  \\\hline
$ N_{tracks}$ & 4--31 & 6--33 & 14--41 \\\hline
$N_{clusters}$ & 10--64 & 18--74 & 43--99 \\\hline
$E_{miss}/\sqrt{s}$ & $<$0.41 & $<$0.41 & 0.08--0.43 \\\hline
${\it Prob}(\chi^2_{WW},5)$ & -- & -- & $<$0.1 \\\hline
${\it Sphericity}$ & $<$0.3 & $<$0.52 & 0.05--0.78 \\\hline
$W_{jet1}$ & 0.02--0.4 & 0.04--0.65 & $\geq 0.35$ \\\hline
$W_{jet2}$ & 0.01--0.6 & 0.04--0.75 & $\geq 0.07$ \\\hline
$|M_{jet1}-M_{jet2}|$ (\gev) & $<$25 & -- & -- \\\hline 
$E_{bgo}/\sqrt{s}$ & 0.08--0.57 & 0.11--0.57 & 0.11--0.57\\\hline
$E_{hcal}/\sqrt{s}$ & 0.04--0.42 & 0.02--0.37 & 0.02--0.37\\\hline

* ${\it Thrust}$ & $>$0.95 & 0.91--0.97 & 0.55--0.92 \\\hline
* ln($\ytq$) & $>-8.63$ & $>-6.40$ & $>-4.74$ \\\hline
* sin($\theta_{miss}$) & $>$0.66 & $>$0.47 & $>$0.62 \\\hline
* $E_{vis}/{\sqrt{s}}$& 0.61--0.80 & 0.35--0.79 & 0.35--0.79 \\\hline
* $\theta_{acop}$ (rad) & $<$3.07 & $<$3.05 & $<$3. \\\hline
\end{tabular}
\caption{Values of the selection requirements for the 
``4 jets + $\Emiss$'' selections.
These final states follow from neutralino decays via 
$\lambda'$ coupling, as listed in Table~\ref{tab:topologies_lqd}.}
\label{tab:lambdaprime_sel3}
\end{center}
\end{table*}


\begin{table*}
\begin{center}

   \begin{tabular}{|l|c|c|c|} \hline

        \multicolumn{4}{|c|} {$\lambda'_{ijk}: \;$
 multijets + leptons }\cr\hline

$N_{clusters}$ & \multicolumn{3}{|c|}{$ \geq$ 13} \\\hline
$N_{tracks}$ & \multicolumn{3}{|c|}{$ \geq$ 10} \\\hline
$E_{v12}/E_{vis}$ & \multicolumn{3}{|c|}{$ <$ 0.30} \\\hline
sin($\theta_{T}$) & \multicolumn{3}{|c|}{$ >$0.139} \\\hline
$E_{par}/E_{vis}$ &\multicolumn{3}{|c|}{$ <$0.5} \\\hline
$E_{perp}/E_{vis}$ &\multicolumn{3}{|c|}{$ <$0.2} \\\hline
$N_{jets8}$ &\multicolumn{3}{|c|}{$ \geq$ 3} \\\hline
$E_{vis}/\sqrt{s}$&\multicolumn{3}{|c|}{$ \geq$ 0.7} \\\hline
sin($\theta_{miss}$) & \multicolumn{3}{|c|}{$ >$0.139} \\\hline
$N_{e,\mu}$ & \multicolumn{3}{|c|}{$ \geq$ 2, 
one isolated ($E^{\ell}_{20} < 1$ \gev, $E_{lepton}> 5$ \gev)} \\\hline
$\mcha$ \rpvpham (\gev) & 50 & 60 & $\geq $ 70 \\\hline
* ${\it Thrust}$ & 0.82--0.89 & 0.74--0.88 &0.51--0.79 \\\hline
* ln($\ytq$) & $>-5.56$ & $>-5.22$ & $> -5.80$ \\\hline
* ln($\yqc$) & $>-5.97$ & $>-5.66$ & $> -5.24$ \\\hline
\end{tabular}
\caption{Cut values for $\lambda'_{ijk}$ chargino 
selections with a multijet topology with at least 2 leptons and no 
missing energy, as detailed in Table \ref{tab:topologies_lqd}.
Different selection criteria are developed according to the mass values 
of the pair--produced charginos.}
 \label{tab:lambdaprime_sel4}
\end{center}
\end{table*}
 
\begin{table*}
\begin{center}
\begin{tabular}{|l|c|c|c|} \hline

        \multicolumn{4}{|c|}{$\lambda'_{ijk}: \;$
 multijets + $\Emiss$ }\cr\hline

$N_{clusters}$ & \multicolumn{3}{|c|}{$ \geq$ 13} \\\hline
$N_{tracks}$ & \multicolumn{3}{|c|}{$ \geq$ 10} \\\hline
$E_{v12}/E_{vis}$ & \multicolumn{3}{|c|}{$ <$ 0.30} \\\hline
sin($\theta_{T}$) & \multicolumn{3}{|c|}{$ >$0.139} \\\hline
$E_{par}/E_{vis}$ &\multicolumn{3}{|c|}{$ <$0.5} \\\hline
$E_{perp}/E_{vis}$ &\multicolumn{3}{|c|}{$ <$0.2} \\\hline
$N_{jets8}$ &\multicolumn{3}{|c|}{$ \geq$ 3} \\\hline
$E_{vis}/\sqrt{s}$&\multicolumn{3}{|c|}{0.5--0.9} \\\hline
sin($\theta_{miss}$) & \multicolumn{3}{|c|}{$ >$0.436} \\\hline
\multicolumn{4}{|c|}{events with one isolated lepton 
($E_{20}^{\ell} <$ 1 \GeV ) are rejected} \\\hline
$\mcha$ \rpvpham (\gev) & 50 & 60 & $\geq $ 70 \\\hline
* ${\it Thrust}$ & 0.82--0.90 & 0.71--0.84 &0.54--0.80 \\\hline
* ln($\ytq$) & $>-5.68$ & $>-5.70$ & $>-4.56$ \\\hline
* ln($\yqc$) & $>-6.17$ & $>-5.94$ & $>-5.21$ \\\hline
* $p^T_{miss}$ & $>$ 10 \GeV & $>$ 20 \GeV & $>$ 20  \GeV\\\hline
\end{tabular}
\caption{Selection criteria for $\lambda'_{ijk}$ chargino 
selections with a multijet topology, missing energy and no isolated 
leptons, as presented in Table \ref{tab:topologies_lqd}.
Different selection criteria are developed according to the mass values of the pair--produced charginos.}
 \label{tab:lambdaprime_sel5}
\end{center}
\end{table*}

\begin{table*}
\begin{center}
\begin{tabular}{|l|c|}
\hline
\multicolumn{2}{|c|}{$\lambda'_{ijk}: \;$
 multijets + lepton(s) and $\Emiss$ }\\\hline
 $N_{clusters}$ & $ \geq$ 13 \\\hline
 $N_{tracks}$ & $ \geq$ 10 \\\hline
$E_{v12}/E_{vis}$ & $ <$ 0.30 \\\hline
sin($\theta_{T}$) &$ >$0.139 \\\hline
$E_{par}/E_{vis}$ & $ <$0.5 \\\hline
$E_{perp}/E_{vis}$ & $ <$0.2 \\\hline
$N_{jets8}$ & $ \geq$ 3 \\\hline
$E_{vis}/\sqrt{s}$ & $<$0.9 \\\hline
sin($\theta_{miss}$) & $ >$0.436 \\\hline
$N_{e,\mu}$ & $ \geq$ 1, one isolated ($E_{20}^{\ell} <$
 1 \GeV, $E_{lepton}<$ 30 \GeV) \\\hline
W mass window & reject if 70 \gev $< M_{qq}<$  90 \gev  \\
              & and 70 \gev $< M_{\ell\nu} <$ 90 \gev  \\\hline
* ${\it Thrust}$ & 0.56--0.87 \\\hline
* ln($y_{34}$) & $>-5.35$ \\\hline
* ln($y_{45}$) & $>-6.22$ \\\hline
* $p^T_{miss}$ & $> 12.3$ \GeV \\\hline
\end{tabular}
\caption{Values of the selection requirements
 for $\lambda'_{ijk}$ chargino 
selections with a multijet topology with both lepton(s) and missing energy.}
 \label{tab:lambdaprime_sel6}
\end{center}
\end{table*}

\begin{table*}
  \begin{center}
\begin{tabular}{|l|c|}  \hline
\multicolumn{2}{|c|}{$\lambda'_{ijk}: \;$
  scalar electrons } \cr \hline

 $N_{clusters}$ &  {$\geq$ 13} \\ \hline
 $N_{tracks}$   &  {$\geq$ 1 } \\ \hline
 ${E_{vis}}/{\sqrt{s}}$ &  {$>$ 0.70 } \\ \hline
 ${E_{perp}}/{E_{vis}}$ &  {$<$ 0.20 } \\ \hline
 ${E_{par}}/{E_{vis}}$  &  {$<$ 0.20 } \\ \hline
 ${E_{v12}}/{E_{vis}}$  &  {$<$ 0.30 } \\ \hline
 ${E_{bgo}}/{E_{vis}}$  &  {0.05 -- 0.98 } \\ \hline
 ${\rtsp}/{\rts}$       &  {$>$ 0.80 } \\ \hline
 sin$(\theta_{T})$      & {$>$ 0.139 } \\ \hline   
 ${\it Thrust}$         & $<$ 0.95 \\ \hline
 ln($\ytq$)             & $> -8.0$ \\ \hline
 ln($\yqc$)             & $> -9.0$ \\ \hline
 $N_{e}$                & $\geq 1 $  \\ \hline   
 $E_1^{\ell}$ (\gev)    & $>$ 5      \\ \hline

\end{tabular}
\caption{Selection criteria of the scalar electron selections, 
for final states
with 4 jets, 2 to 4 leptons and possible missing energy, as shown in
Table \ref{tab:topologies_lqd}.}
 \label{tab:lambdaprime_sel7}
\end{center}
\end{table*}

\begin{table*} 
  \begin{center}
  \begin{tabular}{|l|c|c|c|c|c|} \hline 

    \multicolumn{6}{|c|} {$\lambda''_{ijk}: \;$
    multijets } \cr \hline

 $N_{clusters}$ & \multicolumn{5}{|c|} {$\geq$ 13} \\ \hline
 $N_{tracks}$   & \multicolumn{5}{|c|} {$\geq$ 1 } \\ \hline
 ${E_{vis}}/{\sqrt{s}}$ & \multicolumn{5}{|c|} {$>$ 0.70 } \\ \hline
 ${E_{perp}}/{E_{vis}}$ & \multicolumn{5}{|c|} {$<$ 0.20 } \\ \hline
 ${E_{par}}/{E_{vis}}$  & \multicolumn{5}{|c|} {$<$ 0.20 } \\ \hline
 ${E_{v12}}/{E_{vis}}$  & \multicolumn{5}{|c|} {$<$ 0.30 } \\ \hline
 ${E_{bgo}}/{E_{vis}}$  & \multicolumn{5}{|c|} {0.05 -- 0.98 } \\ \hline
 ${\rtsp}/{\rts}$       & \multicolumn{5}{|c|} {$>$ 0.80 } \\ \hline
 sin$(\theta_{T})$          & \multicolumn{5}{|c|} {$>$ 0.139 } \\ \hline   
    \mch (\gev) & 20 -- 30 & 30 -- 40 & 40 -- 50 & 50 -- 60 & $\geq$ 60  \\ \hline 
 * ${\it Thrust}$   & 0.940 -- 0.973 & 0.901 -- 0.960 & 0.828 -- 0.919 & 0.755 -- 0.889 
                & 0.575 -- 0.844  \\ \hline
 * ln($\ytq$)   & $> -5.85$ & $> -5.70$ & $> -5.20$ & $> -5.16$ & $> -4.77$ \\ \hline
 * ln($\yqc$)   & $> -9.68$ & $> -7.02$ & $> -6.16$ & $> -5.82$ & $> -4.62$ \\ \hline
 * $W_{jet1}$   & 0.12 -- 0.22 & 0.15 -- 0.45 & 0.25 -- 0.55 & 0.21 -- 0.67 
                & 0.38 -- 0.90 \\ \hline

  \end{tabular}                     
  \caption{Cut values of the neutralino selections, for final
   states with at least six hadronic jets. The same multijet topologies
   arise from indirect neutralino and chargino hadronic decays,
   as presented in Table~\ref{tab:topologies_udd}. 
   Different selection criteria are developed according to the
   mass values of the pair-produced supersymmetric particles. }
  \label{tab:lambdasec_sel1}
  \end{center}
\end{table*}

\begin{table*}
\begin{center}
\begin{tabular}{|l|c|c|}\hline

        \multicolumn{3}{|c|} {$\lambda''_{ijk}: \;$
 multijets + lepton(s) }\cr\hline

$N_{clusters}$ & \multicolumn{2}{|c|}{$ \geq$ 13} \\ \hline
$E_{v12}/E_{vis}$ & \multicolumn{2}{c|}{$ <$ 0.20} \\ \hline
$E_{bgo}/E_{vis}$ & \multicolumn{2}{c|}{0.05--0.98} \\ \hline
sin($\theta_{T}$) & \multicolumn{2}{c|}{$ >$0.139} \\ \hline
   & semileptonic & leptonic \\\hline
$E_{par}/E_{vis}$ & $<0.3$ & $<0.5$ \\\hline
$E_{perp}/E_{vis}$ & $<0.3$ & $<0.5$ \\\hline
$ N_{tracks}$ & $>20$ & 10--30 \\\hline
$N_{e,\mu,\tau}$ & $\geq 1$ & $\geq 2$ \\\hline
$E_{20}^{\ell}$ & $> 1$ \GeV & -- \\\hline

* ${\it Thrust}$ & 0.52--0.83 & 0.55--0.79 \\\hline
* ln($\ytq$) & $>-4.55 $&$>-4.15$ \\\hline
* ln($\yqc$) & $>-5.02$ & $>-4.58$ \\\hline
* sin($\theta_{miss}$) & $>0.24$ & $>0.51$ \\\hline
* $E_{vis}/{\sqrt{s}}$& $>0.6$ & 0.52--0.88 \\\hline
\end{tabular}
\caption{Details of final selections for semileptonic and leptonic decays channels of the charginos for the $\lambda''$ coupling, as listed in Table \ref{tab:topologies_udd}.}
\label{tab:lambdasec_sel2}
\end{center}
\end{table*}

\begin{table*}[ht]
  \begin{center}
        \begin{tabular}{|l|l|} \hline

\multicolumn{2}{|c|}{$\lambda''_{ijk}: \;$
  scalar leptons}\cr\hline

$\serr$   &  2 electrons. 
             For at least one of them: $E_{20}^{\ell} > 1$ \GeV \\ \hline
$\smur$   &  2 muons. 
             For at least one of them: $E_{20}^{\ell} > 1$ \GeV \\ \hline
$\staur$  &  {1 lepton (e, $\mu$, $\tau$)} \\ \hline
$\staur$ with high $\Delta M$ & {1 lepton (e, $\mu$, $\tau$)} and 
 $M_{5C} = M_{\tilde{\tau}_R} \pm 5$ \GeV.\\\hline
\end{tabular}
\caption{Cut values, in addition to the ``multijets'' requirements,
 of the scalar lepton selections (Table~\ref{tab:lambdasec_sel1}), for final states
with 6 jets and 2 leptons, as shown in
Table \ref{tab:topologies_udd}.
Different selection criteria are developed according to the 
lepton flavour expected in the final state.}
 \label{tab:lambdasec_sel3}
\end{center}
\end{table*}

\end{document}